\documentclass[twocolumn]{article} 
\usepackage{cite}
\usepackage[utf8]{inputenc} 

\usepackage{pdflscape}
\usepackage{afterpage}
\usepackage{capt-of}

\usepackage{geometry} 
\usepackage{graphicx} 

\usepackage{soul}
\usepackage{siunitx}
\usepackage{booktabs} 
\usepackage[table,xcdraw]{xcolor}
\usepackage{array} 
\usepackage{paralist} 
\usepackage{verbatim} 
\usepackage{subfig} 
\usepackage[nottoc,notlof,notlot]{tocbibind} 

\usepackage{hyperref}
\usepackage{kantlipsum}
\usepackage{eucal}
\usepackage{amsmath,amssymb,amsthm}
\usepackage{mathtools}
\usepackage{enumerate}  
\usepackage{float}
\usepackage{xargs}  
\usepackage[ampersand]{easylist}
\usepackage{soul} 

\usepackage{xcolor}
\usepackage{framed}
\definecolor{shadecolor}{RGB}{242,242,242}
\definecolor{superfluidcolor}{RGB}{211,211,211}

\usepackage{colortbl}
\usepackage{titlesec} 
\renewcommand\thesection{\Roman{section}} 
\renewcommand\thesubsection{\Roman{subsection}} 
\titleformat{\section}[block]{\bfseries}{\thesection.}{1em}{} 
\titleformat{\subsection}[block]{\bfseries}{\thesubsection.}{1em}{} 

\usepackage{authblk}
\usepackage{setspace}
\usepackage{wrapfig}
\usepackage{sidecap}
\usepackage[export]{adjustbox}

\newcommand{\ada}{{a}^{\dagger} {a}}
\newcommand{\ad}{{a}^{\dagger}}

\newcommand{\rhosf}{\rho}
\newcommand{\rhohe}{\rho_{\text{He}}}

\newcommand{\om}{\Omega_{\text{m}}}
\newcommand{\domo}{\delta\Omega}
\newcommand{\azorder}{\mu}
\newcommand{\radnodes}{\nu}
\newcommand{\amp}{\eta_{}}

\newcommand{\phif}{{{\phi}^{(4)}_{\azorder , \radnodes}}}
\newcommand{\phit}{{{\phi}^{(3)}_{\azorder , \radnodes}}}

\newcommand{\avdw}{a_{\text{vdw}}}
\newcommand{\xzpf}{x_{\text{zpf}}}
\newcommand{\xc}{x_{\text{crit}}}
\newcommand{\meff}{m_{\text{eff}}}
\newcommand{\alef}{\alpha_{\text{eff}}}

\usepackage{abstract} 
\title{\textbf{Extreme quantum nonlinearity in superfluid thin-film surface waves}}

\author[1]{Y. L. Sfendla\textsuperscript{*}}
\author[1]{C. G. Baker}
\author[1]{G. I. Harris}
\author[2]{L. Tian}
\author[1]{R. A. Harrison}
\author[1]{W. P. Bowen}
\affil[ ]{$^*$Corresponding author: \textit{y.sfendla@uqconnect.edu.au}}

\affil[1]{ARC Centre of Excellence for Engineered Quantum Systems, School of Mathematics and Physics, The University of Queensland, Brisbane 4072, Australia}
\affil[2]{School of Natural Sciences, University of California, Merced, California 95343, USA}
\date{\small (Dated: 7 November, 2020)}

\begin{document}
\twocolumn[
\vspace{-1.85cm}
\maketitle 
\begin{onecolabstract}

We show that highly confined superfluid films are extremely nonlinear mechanical resonators, offering the prospect to realize a mechanical qubit. Specifically, we consider third-sound surface waves, with nonlinearities introduced by the van der Waals interaction with the substrate. Confining these waves to a disk, we derive analytic expressions for the cubic and quartic nonlinearities and determine the resonance frequency shifts they introduce. We predict single-phonon shifts that are three orders of magnitude larger than in current state-of-the-art nonlinear resonators. Combined with the exquisitely low intrinsic dissipation of superfluid helium and the strongly suppressed acoustic radiation loss in phononic crystal cavities, we predict that this could allow blockade interactions between phonons as well as two-level-system-like behavior. Our work provides a new pathway towards extreme mechanical nonlinearities, and towards quantum devices that use mechanical resonators as qubits. \vspace{1cm}
\end{onecolabstract}
]         
\section*{INTRODUCTION}

Nonlinearities are widely used in quantum technologies. For instance, they allow the generation of nonclassical states \cite{Sletten2019,Arrangoiz-Arriola2019,kues2017,Wang1087,hensen2015,Tian2013}, two-qubit interactions \cite{hacker_photonphoton_2016,debnath2016,arute2019}, and quantum nondemolition measurements \cite{braginsky1980quantum, thorne1978quantum, nakajima2019quantum, lei2016quantum, kono2018quantum}. Sufficiently strong nonlinearities can introduce resolvable anharmonicity in a resonator, so that when resonantly driven it can only absorb a single quantum of energy, mimicking the behavior of a two-level system. This provides the possibility of blockade-type interactions, where phonons (or photons, depending on the resonator) can only pass through the resonator one at a time \cite{dayan2008photon,lemonde_enhanced_2016,Guan_2017}. It also allows artificial qubits to be engineered, such as the superconducting qubits widely used in quantum computing \cite{Buluta_2011}.

Nonlinear \emph{mechanical} resonators have quantum applications ranging from the preparation of nonclassical states \cite{Rips2012,Rips2014,satzinger_quantum_2018,PhysRevLett.115.017202} to quantum-enhanced force sensing \cite{Greywall1994,Babourina-Brooks2008,Woolley2008,szorkovszky2011mechanical,Lu2015}, quantum backaction-evading measurement \cite{Szorkovszky2013}, and mechanical quantum state tomography \cite{Warszawski_2019}. 
Achieving the single-phonon nonlinear regime in a mechanical resonator is of both fundamental and technological importance. It would allow artificial atoms to be built from massive objects consisting of billions of atoms, testing quantum physics in uncharted regimes of macroscopicity, and would provide a new form of qubit for quantum computation among other quantum applications \cite{Guan_2017,Rips}. 

Reaching the single-phonon nonlinear regime in a mechanical resonator requires an intrinsic nonlinearity far stronger than what has been achieved to date \cite{Antoni:2012aa,Lee2002,Maillet:2018aa,Hocke:2014aa,Suh2010,Eichler:2011aa,Huang2016}, combined with exceptionally low dissipation so that the energy level shifts introduced by the nonlinearity are resolvable. Here, we propose to achieve this using a thin spatially confined superfluid helium film, similar to the ones used in recent experimental work on optomechanical cooling \cite{Harris2016}, lasing \cite{he2019strong} and quantized vortex detection \cite{Sachkou2019}. The superfluid resonator is a third-sound surface wave with restoring force provided by the Van der Waals interaction with the substrate. 

\begin{figure*}[t]
\centering
\includegraphics[width=0.95\textwidth]{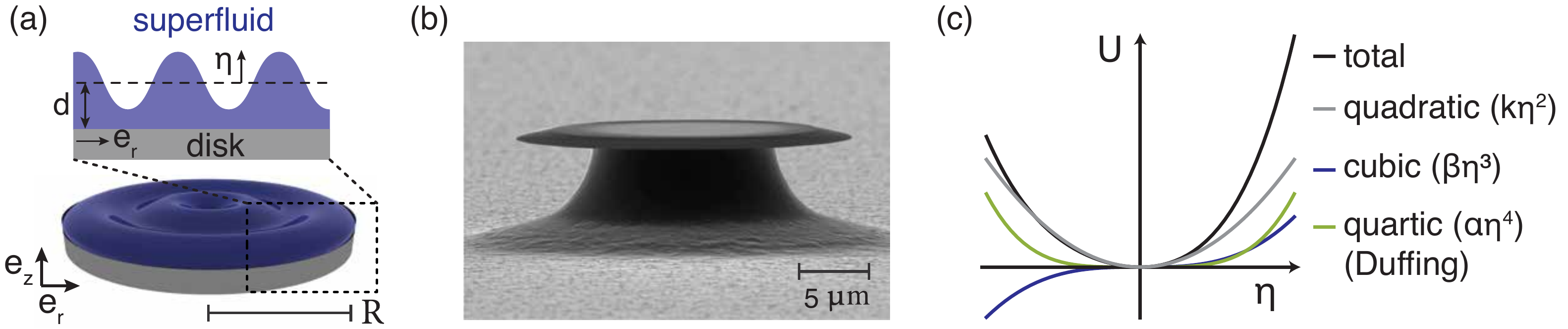}%
 \caption{(a) Illustration of a surface wave of amplitude $\amp[r, \theta]$ in a thin superfluid helium film of mean thickness $d$, confined to a circular geometry of radius $R$. (b) Previously demonstrated experimental methods for circular confinement of a superfluid helium film include adsorption of the film on the surface of an on-chip microdisk as shown in this SEM micrograph \cite{Harris2016,he2019strong,McAuslan2016,Sachkou2019}, or on the inside of parallel disks of a capacitor as in Ref. \cite{Hoffmann:2004aa,Ellis1983}. (c) Anharmonic potential of a superfluid oscillator with spring constant $k$, cubic nonlinear constant $\beta$ and quartic (Duffing) constant $\alpha$.  \label{fig:potential}}
\end{figure*}

We derive an analytical model of the cubic and quartic (Duffing) nonlinearities due to van der Waals forces for a film confined on a circular disk. We find that the nonlinearities depend strongly on the radius of confinement, and predict that the quartic nonlinearity in a \SI{5}{\nano \meter} thick film with \SI{100}{\nano \meter} radius would manifest single-phonon frequency shifts three orders of magnitude larger than those seen in state-of-the-art nonlinear mechanical resonators including graphene sheets \cite{Eichler:2011aa}, carbon nanotubes \cite{Eichler:2011aa} and molecule-coupled resonators \cite{Huang2016}. Our analysis shows that the primary effect of the cubic nonlinearity is to modify the magnitude of the quartic term, consistent with previous work on classical mechanical resonators \cite{Lifshitz2009,nayfeh2008nonlinear,Kozinsky2006}. This result has broad relevance, since a wide variety of mechanical resonators exhibit both cubic and quartic nonlinearities \cite{Lifshitz2009}.

Even with extremely high nonlinearities, achieving sufficiently low dissipation to enter the single-phonon nonlinear regime is a significant challenge. Superfluid helium affords exceptionally low intrinsic dissipation. Indeed, sub-millihertz dissipation rates have been observed in third sound resonators with millimeter dimensions \cite{PhysRevB.39.2703}. However, radiative dissipation increases with increasing spatial confinement.
We introduce the concept of superfluid thin-film phononic crystal cavities to overcome this, showing that radiative dissipation can be greatly suppressed,
even at hundred-nanometer size scales. We further predict that dissipation due to thermalisation and vortices should not be a barrier to reaching the single-phonon nonlinear regime. Together with the level of nonlinearity predicted by our model, this suggests that the single-phonon nonlinear regime can be reached, opening a path to probe quantum macroscopicity in a new domain and to build a new class of qubits for quantum computing and metrology.

\section*{RESULTS}

\subsection*{The anharmonic superfluid oscillator potential}
 \label{sec:anharmonic}

Liquid helium exhibits superfluidity below a critical temperature $T_{\lambda}$. It can then be described as an effective mixture of a normal fluid with density $\rho_n$ and a superfluid with density $\rhosf$, with total density $\rho_{\rm He} = \rhosf + \rho_n$~\cite{Atkins1959, tilley1990superfluidity}. At temperatures well below the critical temperature the ratio $\rhosf/\rhohe$ approaches $1$. For instance, for a critical temperature $T_{\lambda}=\SI{2.2}{\kelvin}$, $\rhosf/\rhohe>0.98$ for temperatures beneath 1~K  \cite{Andronikashvili1946}. In this low-temperature limit, superfluid helium has a combination of traits often sought after in mechanical resonators: low mechanical dissipation arising from near-zero viscosity, and ultralow optical absorption. Indeed, it has been used as the mechanical resonator in several recent optomechanical platforms \cite{Lorenzo_2014,McAuslan2016,Childress2017,Harris2016,Shkarin2019,he2019strong,kashkanova_optomechanics_2017,Kashkanova2017,Rojas2015,Souris2017} and in experiments that study the physics of quantum fluids \cite{Sachkou2019, Forstner2019}.
In these references, the superfluid fills a cavity \cite{kashkanova_optomechanics_2017,Shkarin2019,Kashkanova2017} or channel \cite{Rojas2015}, is levitated as a droplet \cite{Childress2017}, or spread out on a surface as a few-nanometers-thin film\cite{PhysRevB.39.2703,PhysRevLett.71.1577,Harris2016,he2019strong}. The latter case is investigated here. Due to the thinness of the film, the normal component can be considered viscously clamped to the surface~\cite{Atkins1959}, while the superfluid component exhibits 
thickness fluctuations that resemble shallow water waves, as illustrated in Fig. \ref{fig:potential}a. These waves are named ``third sound'' and are unique to two-dimensional superfluid helium films \cite{PhysRevB.39.2703,PhysRevLett.71.1577,Schechter:1998aa}. 

In this work, we consider the superfluid film to be confined to a circular surface of radius $R$. This geometry is quite general, and can be realized for instance by condensing the film on the surface of a microscopic silica disk (see Fig. \ref{fig:potential}b) \cite{Baker2016,he2019strong,McAuslan2016,Harris2016}. That design is attractive because it allows laser light to circulate in the disk. These ``whispering-gallery'' light waves interact strongly with the third-sound waves in the superfluid, and can serve as a tool to observe and control the superfluid motion \cite{Harris2016}. In this study however, we focus on the film's dynamics: while constrained here to a circular disk, we expect our predictions to be qualitatively mirrored in other superfluid thin-film geometries.

A helium atom at height $z$ is attracted to the substrate atoms via the van der Waals force\cite{tilley1990superfluidity,Atkins1959}.
This leads to a height-dependent potential energy per unit mass stored in a film. The analysis here is restricted to films with mean thicknesses $d$ between $1$~and~\SI{30}{\nano \meter}, and with diameters much larger than their thickness.  In this case the potential is well approximated by \cite{PhysRevLett.30.1122,PhysRevA.7.790}
\begin{equation}
V\left[z\right]=-\frac{\avdw }{z^3}\, ,
\label{eq:Vvdw}
\end{equation}
with $\avdw $ the substrate-dependent van der Waals coefficient characterizing the attraction strength. The scaling with height is modified for thicker and lower-aspect-ratio films \cite{Anderson1970,tilley1990superfluidity}; while for thinner films (on the order of a few atomic layers) corrections due to the approximately one inactive atomic layer must be taken into account \cite{PhysRevLett.32.147, Baierlein1997, PhysRevB.18.2155}. The potential provides a restoring force for fluctuations of the film surface.
Turning to Fig. \ref{fig:potential}a, the circularly confined film somewhat resembles a drumhead---and in fact, the helium surface undulates like the skin of a resonating drum. While it is clear that Eq. (\ref{eq:Vvdw}) is nonlinear and therefore does not describe a Hookean potential, in the small amplitude limit (where nonlinearities can be neglected) the resonances of the surface can be described by Bessel modes\cite{Baker2016}. These eigenmodes are, strictly speaking, valid only for a linear oscillator. However, they are a good approximation for the high quality (see Results) mechanical resonances considered here where the nonlinearity only shifts the mechanical resonance frequency by a small fraction. 

\begin{figure}[h]
\centering
\includegraphics[width=0.40\textwidth]{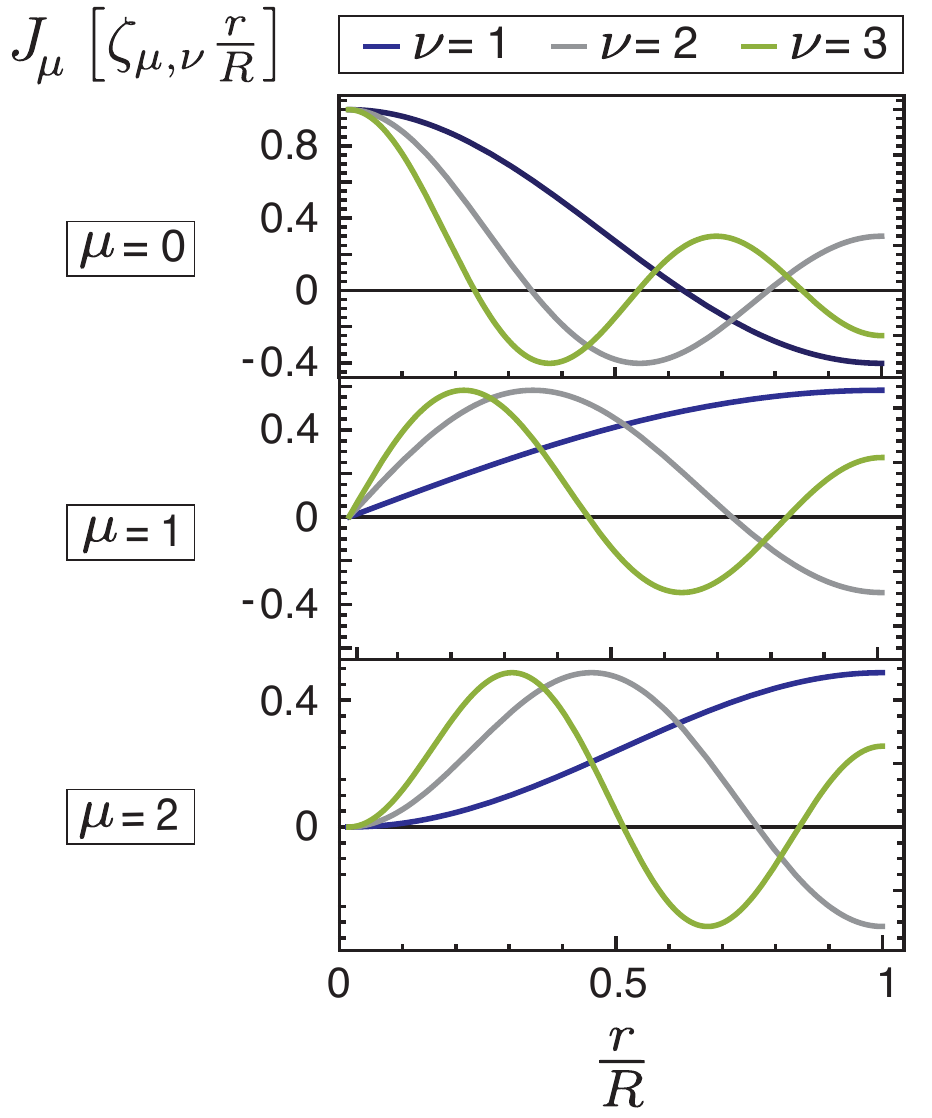}%
 \caption{Bessel modes with free boundary conditions $J_\azorder\left[\zeta_{\azorder , \radnodes} \frac{r}{R}\right]$ of (azimuthal) orders $\azorder=0$, 1 and 2, for radial mode number $\radnodes = 1$ to 3.\label{fig:besselmodes} \vspace*{-0cm}}
\end{figure}

The time-dependent Bessel mode amplitude $h\left[r,\theta, t\right]$ that quantifies the deviation of the film height from the mean thickness $d$ with polar coordinates $r$ and $\theta$ is given by \cite{Baker2016} 
\begin{equation}
h\left[r,\theta, t\right]= \amp \left[r,\theta \right] \sin\left(\om t\right)
\label{eq:besselt}
\end{equation}
with
\begin{equation}
\amp \left[r,\theta \right]= A \,  J_\azorder\left[\zeta_{\azorder , \radnodes} \frac{r}{R}\right]\cos\left( \azorder  \theta\right) .
\label{eq:bessel}
\end{equation}
Here, $A$ is the amplitude, $J_{\azorder}$ is the Bessel function of the first kind of integer order $ \azorder $, and $\om =(\zeta_{\azorder , \radnodes} c_3)/R$ is the mechanical resonance frequency in the absence of nonlinearity. Here the resonance frequency depends on the superfluid speed of sound $c_3=\sqrt{3 \avdw (\rhosf/ \rhohe) \, d^{-3}}$ and a parameter $\zeta_{\azorder , \radnodes}$, which depends on the boundary conditions. In the absence of flow across the resonator boundary, the film is described by \emph{volume-conserving} Bessel modes, i.e., Bessel functions with free boundary conditions \cite{Baker2016}. These mode amplitudes have an extremum at $r=R$: a condition met by choosing $\zeta_{\azorder , \radnodes}$ as the $\radnodes^{\mathrm{th}}$ zero of $J_{\azorder}^{'}$. We restrict our analysis to these boundary conditions since they match previous observations with disk resonators in Ref. \cite{Sachkou2019}. However, fixed boundary conditions could alternatively be used by choosing the zeroes of $J_{\azorder}$ for the coefficient $\zeta_{\azorder , \radnodes}$.
$\amp[r, \theta]$ is the time-independent amplitude of a drumhead mode typically specified by its mode numbers $(\azorder, \radnodes)$ with the order $\azorder$ the number of nodal diameters, also called \emph{azimuthal mode number} (i.e.,  $2 \azorder$ is the number of zeros in the azimuthal direction for $\theta$ = 0 to $2 \pi $) and $\radnodes$ the number of nodal circles, also called \emph{radial mode number} (i.e., $\radnodes$ is the number of zeros in the radial direction for $r=0$ to $R$). The bessel mode function $J_\azorder\left[\zeta_{\azorder , \radnodes} \frac{r}{R}\right]$ is graphed in Fig. \ref{fig:besselmodes}, and the values $\zeta_{\azorder , \radnodes}$ for the first three mode numbers are tabulated in Table \ref{tab:zeta}. 

\begin{table}[t]
\centering
\begin{tabular}{@{}lrrr@{}}
\toprule
\multicolumn{4}{c}{$\zeta_{\azorder , \radnodes}$} \\ \midrule
 & \textit{$\radnodes =$ 1} & \textit{$\radnodes =$ 2} & \textit{$\radnodes =$ 3} \\
\textit{ $\azorder$ =0} & 3.83 & 7.02 & 10.2 \\
\textit{ $\azorder$ =1} & 1.84 & 5.33 & 8.54 \\
\textit{ $\azorder$ =2} & 3.05 & 6.71 & 9.97 \\ \bottomrule
\end{tabular}
\caption{The first three roots ($\radnodes = 1$ to 3) of the derivative of the Bessel function  $J_{\azorder}^{'}$ \cite{Olver:1960aa}.}
\label{tab:zeta}
\end{table}

The van der Waals potential energy stored in the surface deformation $U$ is obtained by integrating the potential $V$ over the deviation from equilibrium \cite{Baker2016}:
\begin{equation}
\begin{aligned}
 U  &=\rhosf\int_{0}^{2\pi}\int_{0}^R\int_d^{d+\amp\left[r, \theta\right]} V\left[z\right] \, r \,\mathrm{d}r\, \mathrm{d}\theta \, \mathrm{d}z 
\end{aligned}
\label{eq:Epot1}
\end{equation}
With Eq. (\ref{eq:Vvdw}) and the Taylor series expansion, we obtain
\begin{align}
\int_d^{d+\amp\left[r, \theta\right]} V\left[z\right]\mathrm{d}z = \frac{\avdw}{2} \left(  \frac{1}{(d+\amp\left[r, \theta\right])^2} - \frac{1}{d^2} \right) \\ = \frac{\avdw}{2 d^2} \sum_{j=0}^{\infty} (j+2) \left(  - \frac{\amp\left[r, \theta\right]}{d}  \right) ^{j+1} .
\end{align}
The majority of previous work on third sound considers only the first two terms in this expansion, which describe the potential of an harmonic oscillator. The higher order terms introduce nonlinearities. Experiments have investigated a variety of nonlinearities \cite{Ellis2004,Ellis1998,Baierlein1997,Ellis1993,Wang2000,Hoffmann:2004aa,Schechter:1998aa} in millimeter and centimeter-scale devices, far away from the regime we consider here.

The first term (proportional to $\amp$) averages out to zero for volume-conserving Bessel modes with $\int_r\int_\theta \amp\left[  r, \theta\right]=0$. The third- and fourth-order terms respectively represent cubic and quartic nonlinearities. We neglect terms of fifth order in $\amp$ and higher, expecting them to be small compared to these first two nonlinear terms. Thus, Eq. (\ref{eq:Epot1}) becomes 
\begin{equation} \label{eq:Epotexpanded} 
\begin{split}
 U =\rhosf \int_{0}^{2\pi}\int_0^R \left(\underbrace{\frac{3\, \avdw  \,\amp^2\left[ r, \theta \right]}{2\, d^4} }_{\text{linear spring}} \underbrace{ - \frac{2\,\avdw  \,\amp^3\left[ r, \theta \right]}{ d^5}}_{\text{quadratic spring}}
\right. 
\\
\left. 
 + \underbrace{\frac{5\,\avdw  \,\amp^4\left[ r, \theta \right]}{2\, d^6}}_{\text{cubic spring}}  \right) \, r \,\mathrm{d}r\, \mathrm{d}\theta , \end{split}
\end{equation}
where we have identified the quadratic, cubic and quartic potential energies associated with linear, quadratic and cubic spring terms.\footnote{In this work we label the nonlinearities by the orders in which they appear as energies and potentials, because we work exclusively in the Hamiltonian formalism. In some other work, especially early work on spring forces, they may be labeled by their (lower) order in force and acceleration equations.} By introducing a reference point $x:=\amp[R, 0]$---that is, the displacement at the periphery of the disk at an angular location $\theta=0$---we can rewrite the potential energy as
\begin{equation}
 U =\frac{1}{2}k\,x^2+\frac{1}{3} \beta x^3 + \frac{1}{4} \alpha x^4 .
\label{eq:epot}
\end{equation}
This is the potential energy of an anharmonic oscillator (see Fig. \ref{fig:potential}c) with restoring force $F=-\nabla U=-k\,x-\beta\, x^2 -\alpha\, x^3$, where $k$ is the linear spring constant given by
\begin{equation} \label{eq:k1}
k=\frac{3 \,\rhosf\, \avdw }{d^4} \int_{0}^{2\pi}  \int_0^R \, \frac{\amp^2\left[ r, \theta \right]}{\amp^2\left[ R, 0\right]} \, r \,\mathrm{d}r\, \mathrm{d}\theta ,
\end{equation}
and $\beta$ and $\alpha$ are the nonlinear spring constants. The \emph{cubic nonlinearity} is given by
\begin{equation} \label{eq:beta1}
\beta=-\frac{6 \,\rhosf\, \avdw }{d^5} \int_{0}^{2\pi}   \int_0^R \, \frac{\amp^3\left[ r, \theta \right]}{\amp^3\left[ R, 0\right]}\, r \,\mathrm{d}r\, \mathrm{d}\theta
\end{equation}
and the \emph{quartic} (also known as \emph{Duffing}) \emph{nonlinearity} is given by
\begin{equation} \label{eq:alpha1}
\alpha=\frac{10 \,\rhosf\, \avdw }{d^6} \int_{0}^{2\pi}  \int_0^R \, \frac{\amp^4\left[ r, \theta \right]}{\amp^4\left[ R, 0\right]}\, r \,\mathrm{d}r\, \mathrm{d}\theta .
\end{equation}

By evaluating the integrals (see Supplementary Information), the strong dependence of the (non)linear spring constants on the film thickness $d$ and confinement radius $R$ is exposed:

\begin{align} \label{eq:k}
k &= \left( 1+ \delta_{\azorder 0} \right) 3 \pi \,\rhosf \avdw  \, {\phi}^{(2)}_{\azorder , \radnodes} \, \frac{R^2}{d^4} \\
& = \left( 1+ \delta_{\azorder 0} \right) \frac{3 \pi}{2}  \rhosf \avdw \left( 1- \frac{ \azorder ^2}{\zeta_{\azorder , \radnodes}^2} \right) \, \frac{R^2}{d^4} \, ,
\end{align}

\begin{align} \label{eq:beta}
\beta = - \delta_{\azorder 0} \, 12 \pi \,  \rhosf \avdw   \, {\phi}^{(3)}_{0 , \radnodes} \, \frac{R^2}{d^5} 
\end{align}
and
\begin{align} \label{eq:alpha}
\alpha = \left( 3+ 5 \delta_{\azorder 0} \right) \frac{5 \pi}{2} \,\rhosf \avdw  \, {\phi}^{(4)}_{\azorder , \radnodes} \, \frac{R^2}{d^6} \, .
\end{align}
Here we have introduced the Kronecker delta function $\delta$, and integrals of the Bessel function ${\phi}^{(p)}_{\azorder , \radnodes} :=  \frac{ \int_0^{\zeta_{\azorder , \radnodes}}  J_{\azorder}^{p}\left[q\right] \, q \, \mathrm{d}q}{\zeta_{\azorder , \radnodes}^{2}  J_{\azorder}^{p}\left[\zeta_{\azorder , \radnodes}\right] }$ for $p= \{ 2,3,4\}$ tabulated in Table \ref{tab:int}. It is worth noting that the cubic nonlinearity $\beta$ vanishes for all but the rotationally invariant ($\azorder = 0$) modes.

The film thickness $d$ can be independently determined and tuned in situ, as we demonstrated in previous work \cite{Harris2016}, while the confinement radius $R$ can be changed through choice of device geometry. By controlling these, Eq. (\ref{eq:beta}) and (\ref{eq:alpha}) reveal that it is possible to access a wide range of cubic and quartic nonlinearities.

Although the relationships found here suggest that the nonlinear coefficients are larger for thinner films and larger radii, the desired parameter regime depends on the relative magnitudes of the nonlinear and linear coefficients, and the mass. In the following section, we will derive what exactly that regime is and what platform parameters one should aim for, in order to reach it. 

\begin{table}[h]
\centering
\begin{tabular}{@{}lrrr@{}}
\toprule
\multicolumn{4}{c}{${\phi}^{(2)}_{\azorder , \radnodes}$} \\ \midrule \midrule
 & \textit{$\radnodes =$ 1} & \textit{$\radnodes =$ 2} & \textit{$\radnodes =$ 3} \\
\textit{ $\azorder$ =0} & $ 1/2 $ & $1/2$ & $1/2$ \\
\textit{ $\azorder$ =1} & \num{0.353	} & \num{0.482	} & \num{0.493} \\
\textit{ $\azorder$ =2} & \num{0.286	} & \num{0.456	} & \num{0.480} \\  \midrule
\multicolumn{4}{c}{${\phi}^{(3)}_{\azorder , \radnodes}$} \\ \midrule \midrule
\textit{ $\azorder$ =0} & \num{- 0.437	} & \num{0.259	} & \num{- 0.236} \\
\midrule
\multicolumn{4}{c}{${\phi}^{(4)}_{\azorder , \radnodes}$} \\ \midrule \midrule
\textit{ $\azorder$ =0} & \num{1.28	 } & \num{1.48	} & \num{1.61} \\
\textit{ $\azorder$ =1} & \num{0.290	} & \num{0.837	} & \num{1.03} \\
\textit{ $\azorder$ =2} & \num{0.223	} & \num{0.704	} & \num{0.891} \\ \bottomrule
\end{tabular}
\caption{Coefficients ${\phi}^{(p)}_{\azorder , \radnodes} =  \frac{ \int_0^{\zeta_{\azorder , \radnodes}}  J_{\azorder}^{p}\left[q\right] \, q \, \mathrm{d}q}{\zeta_{\azorder , \radnodes}^{2}  J_{\azorder}^{p}\left[\zeta_{\azorder , \radnodes}\right] }$ for $p= \{ 2,3,4\}$ with $\zeta_{\azorder , \radnodes}$ the $\radnodes^{\text{th}} $ zero of the Bessel function $J_{\azorder}^{'}$. 
}
\label{tab:int}
\end{table}

 \subsection*{Single-phonon transition resonances}
 \begin{figure}[h]
\centering
\includegraphics[width=0.42\textwidth]{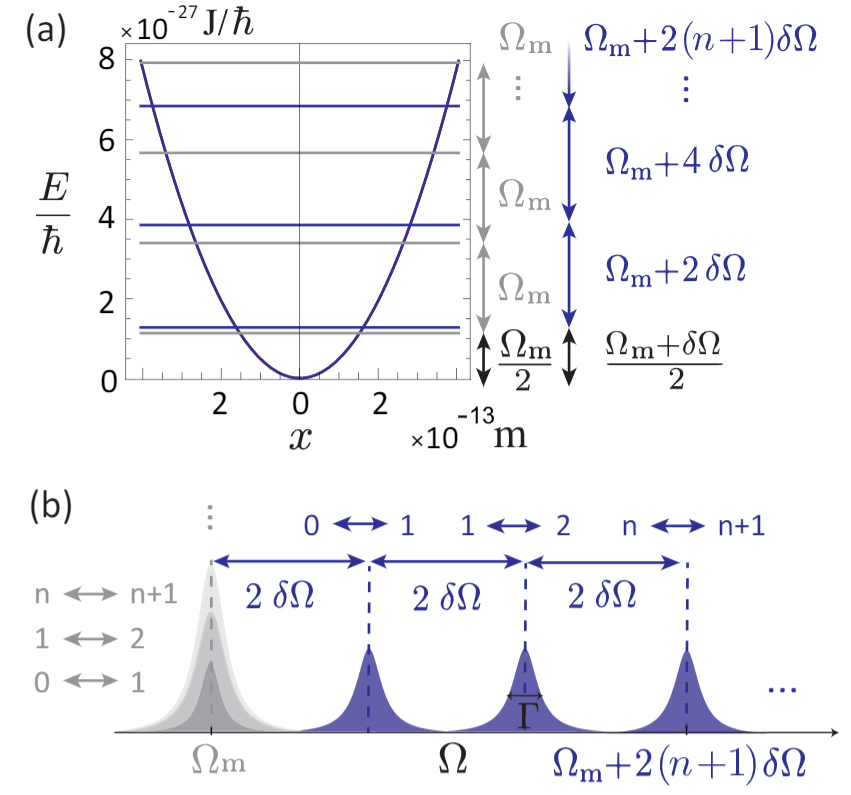}%
 \caption{(a) Energy spectrum (analytical solutions) for $R=\SI{0.5}{\micro \meter}$, $ d=\SI{10}{\nano \meter}$ of a linear (gray) and nonlinear (blue) resonator, with energy shifts enlarged by a factor $10^8$ to facilitate visual representation. (b) Illustration of the corresponding single-phonon transition frequencies. The harmonic oscillator features only a single resonance at $\om$ (gray); the anharmonic oscillator (blue) exhibits degenerate spectral features at $\Omega[n] = \om + (n+1) \domo$.
  \label{fig:spectrumshifts}}
\end{figure} 
 
We consider the Hamiltonian for the oscillator with natural frequency $\om$, effective mass $\meff=k/ \om^2$ and zero-point fluctuation amplitude $\xzpf = \sqrt{\hbar / 2 \meff \, \om}$:
 \begin{align}
H &= \frac{p^2}{2 \meff} + U \label{eq:ham} \\ &= \hbar \om (n+\frac{1}{2}) + \frac{\xzpf^3}{3} \beta (a + \ad)^3+ \frac{\xzpf^4}{4} \alpha (a + \ad)^4 \nonumber
\end{align}
with $a$ ($\ad$) the phonon annihilation (creation) operators satisfying the commutation relation $[a,\ad]=1$, the canonical position and momentum operators $x$ and $p$ satisfying $x = \xzpf \, (a + \ad)$ and $[x,p]=i\hbar$, and $n=\ada$ the phonon number.
The nonlinearities modify the eigenenergies of the oscillator (see Fig. \ref{fig:spectrumshifts}). To second order in perturbation theory we find that the transition between states $n$ and $n+1$ occurs at frequency
\begin{equation} \label{eq:deltaEn}
\Omega[n] = \om + (n+1) \domo 
\end{equation}
(see Supplementary Information) with the single-phonon nonlinear shift
\begin{align}
\domo := 3 \, \frac{\xzpf^4 \, \alef}{\hbar} \label{eq:domo1} 
\end{align}
and $\alef[\alpha, \beta]$ the effective quartic nonlinearity absorbing the cubic nonlinearity:
\begin{equation}
\alef = \alpha - \frac{20}{9} \frac{\xzpf^2 \beta^2}{\hbar \om}  = \alpha - \frac{10}{9} \frac{\beta^2}{k} .
\label{eq:effduff}
\end{equation}
The frequency shift in Eq. (\ref{eq:domo1}) is the mechanical analog of the Kerr nonlinearity in quantum electrodynamics \cite{PhysRevLett.79.1467}, but with the quartic nonlinear coefficient $\alpha$ reduced to $\alef$ due to the presence of the cubic nonlinearity. The modification, Eq. (\ref{eq:effduff}), agrees with the classical result found in Ref.\cite{Lifshitz2009,nayfeh2008nonlinear,Kozinsky2006}.

When the mechanical decay rate (or linewidth) $\Gamma$ of the oscillator is smaller than the spectral splitting, i.e.,
\begin{equation}
\Gamma < \domo , \label{eq:domoquantumcondition}
\end{equation} 
the resonator is sufficiently anharmonic that a single absorbed phonon shifts the frequency off resonance for phonons that arrive later, causing phonon blockade: it behaves more like a two-level system than an harmonic oscillator \cite{Sletten2019,Arrangoiz-Arriola2019}.

The criterion in Eq. (\ref{eq:domoquantumcondition}) can be re-expressed in terms of the critical amplitude $\xc = \sqrt{\frac{2}{3} \frac{ \meff \,  \Gamma \om}{\alpha}}$, which defines the amplitude at which a classical Duffing resonator become bistable. In this form the criterion is $\xzpf > \xc$, that is, the mechanical zero-point motion must exceed the critical amplitude to reach the single-phonon nonlinear regime \cite{Huang2016}.

Substituting the superfluid spring constants from Eq. (\ref{eq:k}) to (\ref{eq:alpha}) yields
\begin{align}
\frac{\alef }{\alpha} = 1-  \delta_{\azorder 0} \frac{8}{3} \frac{\left( {\phi}^{(3)}_{0 , \radnodes} \right)^2}{{\phi}^{(4)}_{0 , \radnodes} } .
\label{eq:effduffexpanded}
\end{align}
Hence, the effective modification of the Duffing nonlinearity by the cubic nonlinearity $\beta$ depends only on the mode symmetry numbers $\azorder$ and $\radnodes$.  Furthermore, as $\radnodes$ increases, $\alef$ rapidly approaches $\alpha$. Indeed, since
\begin{align}
\lim_{\radnodes\to\infty} \frac{\left( {\phi}^{(3)}_{0 , \radnodes} \right)^2}{{\phi}^{(4)}_{0 , \radnodes} } = \frac{0}{\infty} = 0 , \nonumber
\end{align}
we have
\begin{align}
\lim_{\radnodes\to\infty} \frac{\alef }{\alpha} = 1 .
\end{align}
For example, for the $(\azorder=0$; $\radnodes = 1)$ mode, $\alef = 0.6 \, \alpha$ while for a $(\azorder=0$; $\radnodes = 10)$ mode $\alef = 0.98 \, \alpha$.

\subsection*{Open quantum system spectrum}

\begin{figure}[H]
\includegraphics[width=0.47\textwidth]{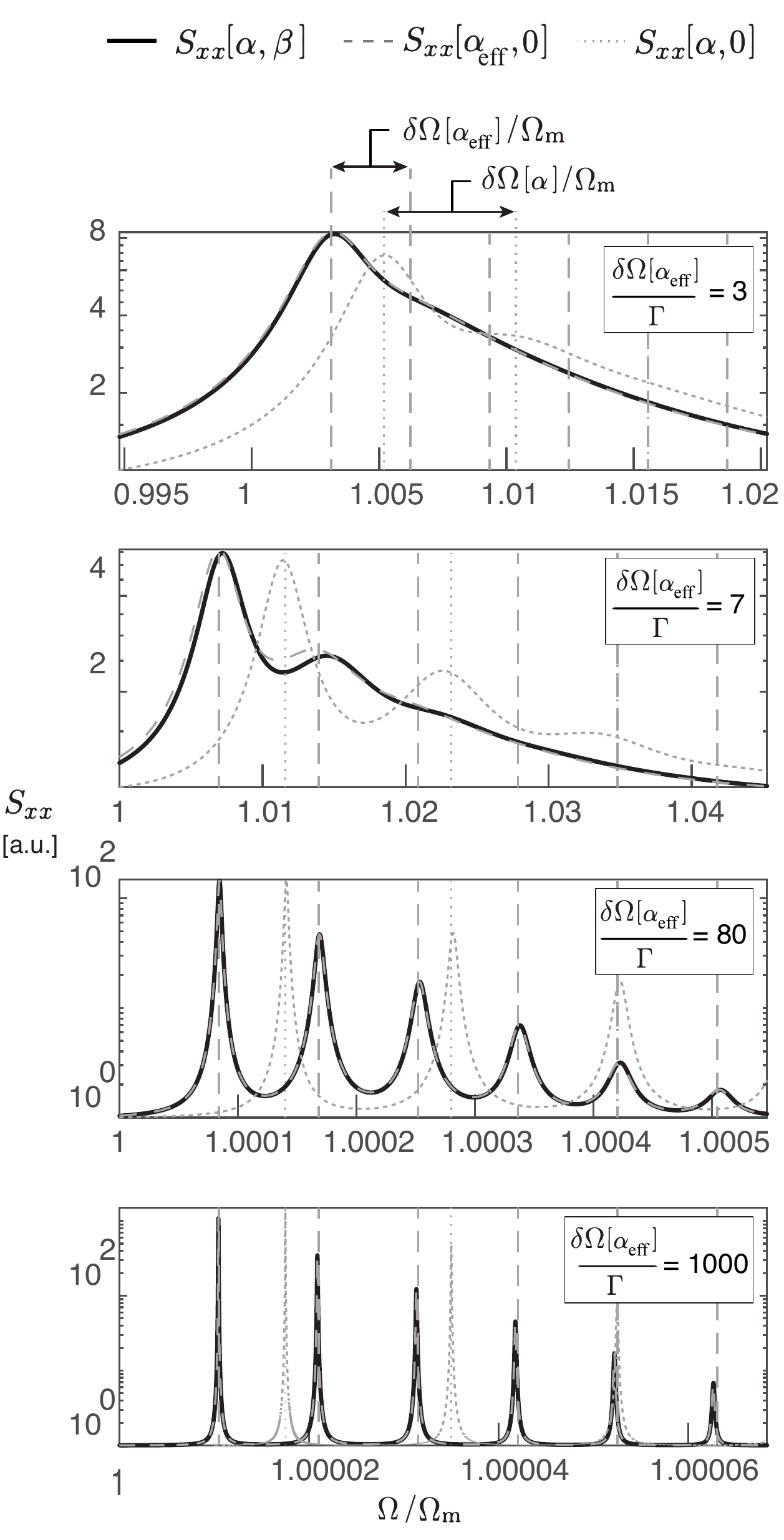}  \caption{Numerically simulated spectra for the full master equation with a quartic Duffing ($\alpha$) and cubic ($\beta$) nonlinearity of the potential ($S_{xx}[\alpha,\beta]$, solid lines), the approximated model where the cubic nonlinearity manifests as a correction to the quartic Duffing strength ($S_{xx}[\alef,0]$, dashed lines), and the spectra in absence of a cubic nonlinearity ($S_{xx}[\alpha,0]$, dotted lines). Vertical lines at $ (E_{n+1} - E_{n})/ \hbar$ for $n= 0,1,2, \,  \dots$ indicate the analytic values of transition energies according to Eq. (\ref{eq:deltaEn}) with $\domo[\alef]$ (dashed) and $\domo[\alpha]$ (dotted). Input parameters: $ T = \SI{50}{\milli \kelvin}$, and from top to bottom:  $R= \SI{3}{\nano \meter}$, $d= \SI{5}{\nano \meter}$, $Q=\num{1E3}$; $R= \SI{3}{\nano \meter}$, $d= \SI{25}{\nano \meter}$, $Q=\num{1E3}$; $R= \SI{10}{\nano \meter}$, $d= \SI{5}{\nano \meter}$, $Q=\num{1E6}$; $R= \SI{20}{\nano \meter}$, $d= \SI{5}{\nano \meter}$, $Q=\num{1E8}$. 
 \label{fig:lindblad_spectra}}
\end{figure}

To determine the expected resonator spectrum in the presence of decoherence and validate our perturbation theory calculation, we numerically solve the Lindblad master equation for a resonator with cubic and quartic nonlinearities (see Supplementary Information). We compare the resulting correlation spectra $S_{xx}$ for three cases: a superfluid resonator with finite quartic and cubic nonlinearity $S_{xx}[\alpha,\beta]$, a pure Duffing resonator $S_{xx}[\alpha,0]$, and a pure Duffing resonator with an effective quartic strength $S_{xx}[\alef,0]$ conform to Eq. (\ref{eq:effduffexpanded}). They are shown, for a range of single-photon nonlinear strengths $\domo / \Gamma$, in Fig. \ref{fig:lindblad_spectra}. 

It is evident at a glance that as the single-phonon nonlinear strength $\domo / \Gamma$ exceeds unity, the nonlinearity lifts the oscillator's spectral degeneracy. The spectra for pure Duffing resonators $S_{xx}[\alpha,0]$ manifest transition resonances\footnote{The small discrepancy stems from the numerical error on the bare energy eigenstates $E_n$, which increases with $n$. This absolute error is inversely proportional to the size of the basis spanning the Hilbert space in the numerical algorithm; the smaller $\alpha$ and $\domo$, the larger the basis must be chosen to mitigate the relative error on the transition frequencies.}
at the
 values $\Omega[n] = \om + (n+1) \domo[\alpha]$ obtained analytically from first-order perturbation theory. The pure Duffing spectrum does, however, differ significantly from the full-Hamiltonian third-sound resonator spectrum $S_{xx}[\alpha,\beta]$, and from that of the effective Duffing third-sound resonator $S_{xx}[\alef,0]$. The latter two are nearly identical in all cases, both in terms of amplitude and transition resonance frequencies. This indicates that the third-sound resonator with a quadratic nonlinearity $\beta$ is accurately described by the effective Duffing resonator according to Eq. (\ref{eq:effduffexpanded}), and its single-phonon transition frequencies are correctly analytically predicted as $\Omega[n] = \om + (n+1) \domo[\alef]$ from Eq. (\ref{eq:domo1}).

\subsection*{Single-phonon nonlinear shift for a superfluid resonator}

Using Eq. (\ref{eq:k}) to (\ref{eq:alpha}) and (\ref{eq:effduff}) we can express $\domo$ in terms of the adjustable parameters $R$ and $d$. With
\begin{equation} \label{eq:xzpf}
\xzpf=\sqrt{ \frac{\hbar }{ (1+ \delta_{\azorder 0}) \pi \sqrt{3 \avdw } \sqrt{\rhosf \rhohe}  \left( 1- \frac{ \azorder ^2}{\zeta_{\azorder , \radnodes}^2} \right) }} \, \, \zeta_{\azorder , \radnodes}^{\frac{1}{2}}  \, \frac{d^{\frac{5}{4}}}{R^\frac{3}{2}}
\end{equation}
one obtains 

\begin{align}
\domo =  \frac{15 \hbar}{(2 + \delta_{\azorder 0}) \pi \rhohe}  \frac{\phif- \delta_{\azorder 0} \frac{8}{3} \phit^2}{\left( 1- \frac{ \azorder ^2}{\zeta_{\azorder , \radnodes}^2} \right)^2}  \frac{ \zeta_{\azorder , \radnodes}^2 }{R^4 d}. 
\label{eq:domofull}
\end{align}

\begin{figure}[h]
\centering
\includegraphics[width=0.45\textwidth]{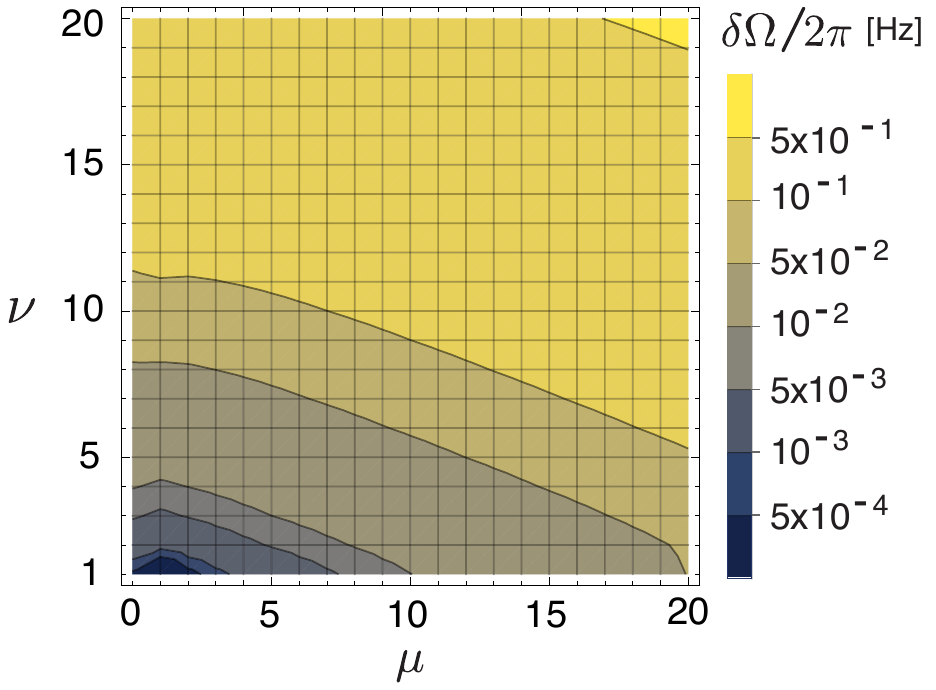}%
 \caption{Resonance frequency shift induced by a single phonon, $\domo = 3 \alef \xzpf^4 \, / \hbar$, as function of the Bessel mode order $\azorder$ and radial node number $\radnodes$ for a \SI{5}{\nano \meter} thick superfluid film confined to a \SI{1}{\micro \meter} radius.  \label{fig:omshift_az_rad}}
\end{figure}

This equation is valid in the limit of a third sound wavelength large compared to the film thickness $d$, and a motional amplitude $\eta \ll d$, which is the case for all examples considered here.” It identifies every parameter available to the researcher who seeks to maximize the single-phonon nonlinear strength of the superfluid resonator: the shift grows strongly with decreasing confinement radius $R$, scales with the inverse of the film thickness $d$, but is independent of the van der Waals coefficient $\avdw$ between the helium film and the substrate. 
The shift is always positive, so the oscillator is effectively ``spring-hardened''. Its dependence on the mode numbers $(\azorder ; \radnodes )$ is rather intricate due to the lack of a closed form of the coefficients $\zeta_{\azorder , \radnodes}$ and ${\phi}^{(4)}_{\azorder , \radnodes} $ (see Supplementary Information). Therefore, we have graphed the frequency shift as a function of the mode numbers $(\azorder ; \radnodes )$ in Fig. \ref{fig:omshift_az_rad}. In this figure and henceforth we take $\rhosf/\rho_{\rm He} \rightarrow 1$, considering only the regime where the temperature is well beneath $T_\lambda$. It can be seen that the nonlinear strength $\domo$ can grow by four orders of magnitude when the Bessel mode order $\azorder$ and radial mode number $\radnodes$ vary from 0 (for the mode order) and 1 (for the radial mode number) to 20.

The predicted nonlinearity of the superfluid film is compared to other systems in Table \ref{tab:overview} for a range of confinement radii $R$ and radial nodes $\radnodes$. From the table it is clear that the \emph{simultaneous} attainment of large $\alpha$ and $\xzpf$ is nontrivial for many platforms, while for the superfluid third-sound resonator both are intrinsically large and tunable. 
When a \SI{5}{\nano \meter} thick superfluid film is confined to a radius of \SI{10}{\micro \meter}, its predicted single-phonon frequency shift surpasses those in membranes\cite{Antoni:2012aa}, cantilevers\cite{Lee2002}, and $\mathrm{Si}_{3}\mathrm{N}_{4}$ beams\cite{Maillet:2018aa,Hocke:2014aa,Suh2010}. For $R =\SI{1}{\micro \meter} $, it exceeds levitated nanoparticles that inherit their strong nonlinearity from an optical trapping potential\cite{Ricci:2017aa} and resonators with an engineered chemical bond\cite{Huang2016}. Finally, in the submicron regime, single-phonon nonlinear shifts might surpass by orders of magnitude those of carbon nanotubes and graphene sheets\cite{Eichler:2011aa}, exceeding the single-phonon nonlinear threshold. 

\subsection*{Reaching the single-phonon nonlinear regime in confined thin superfluid helium}
\label{sec:singlephonon}
 Having obtained the parameter space required to bring a thin-film superfluid resonator into the single-phonon nonlinear regime, the question becomes: Are there platforms available that may facilitate these requirements? Can one reasonably engineer an on-chip superfluid resonator whose damping $\Gamma$ approaches the single-phonon frequency shift $\domo$? 

\begin{figure}[h]
\centering
\includegraphics[width=0.48\textwidth]{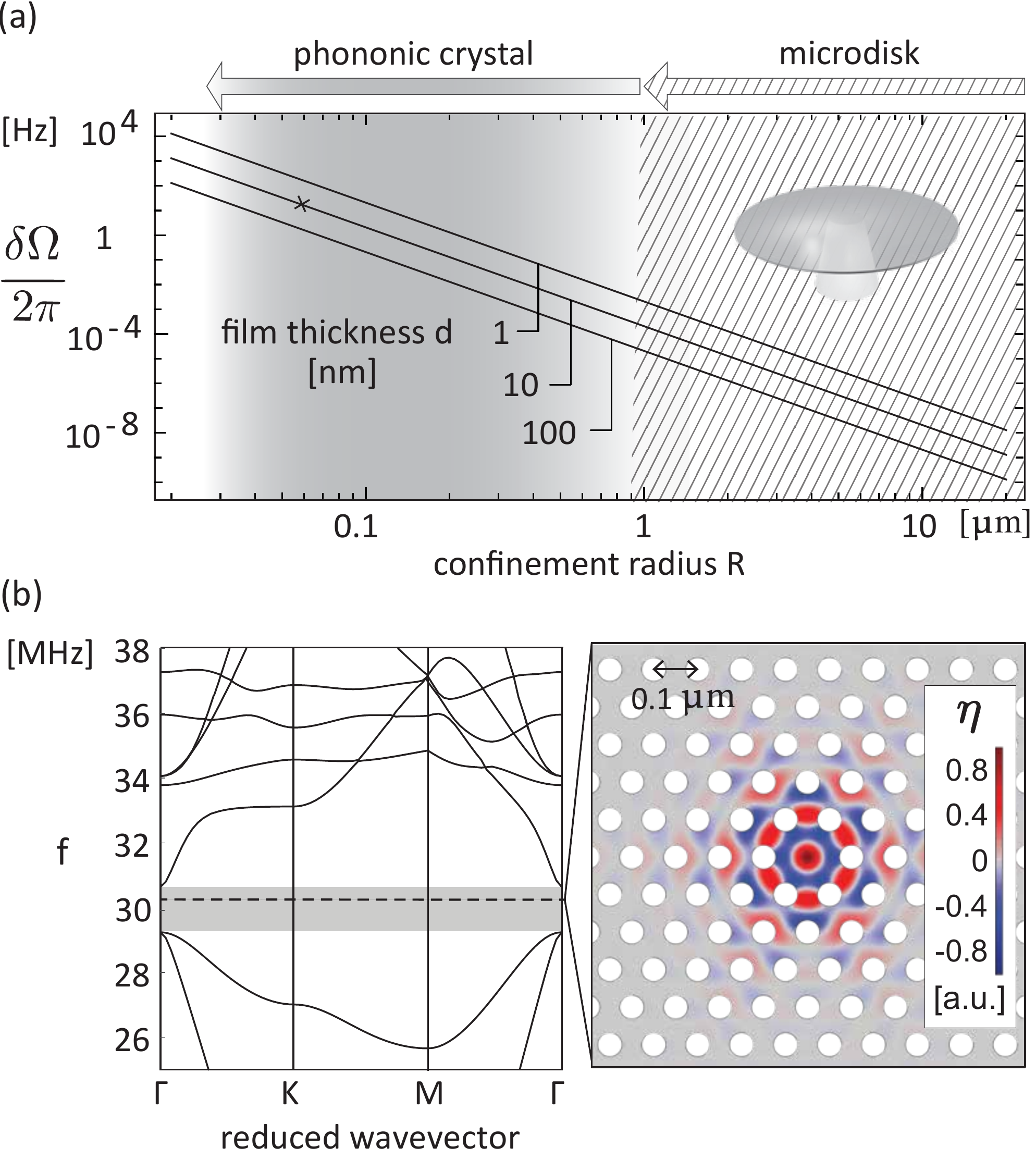}%
 \caption{(a) Resonance frequency shift induced by a single phonon $\domo = 3 \alef \xzpf^4 \, / \hbar$ as a function of mode confinement radius $R$ for various superfluid film thicknesses $d$ for the $( \azorder = 0; \radnodes = 1)$ mode. The quantized nature of the anharmonic oscillator energy spectrum is resolvable if the mechanical linewidth $\Gamma < \domo $. Hatched: confinement radii for microdisks\cite{Harris2016,he2019strong,McAuslan2016,Sachkou2019}. Gray shaded: confinement radii achievable through phononic-crystal (PnC) trapping. X-mark: mode confined in PnC cavity with lattice constant \SI{100}{\nano \metre}. (b) PnC band structure and complete band gap (gray shaded) for a hexagonal honeycomb lattice with holes \SI{55}{\nano \metre} in diameter, lattice constant \SI{100}{\nano \metre} and the van der Waals coefficient for silicon $\avdw = \SI{3.5E-24}{m^5 . s^{-2}} $ \cite{PhysRevLett.30.1122}. A central defect (absence of hole) introduces a Bessel-like mode at \SI{30}{\mega\hertz}---within the band gap---as shown by the finite-element-method simulation. \label{fig:omshift_r_d_band_structure}}
\end{figure}

 In the earliest third-sound resonators \cite{Ellis1983,Hoffmann:2004aa}, the superfluid film was adsorbed on the inside of two parallel metalized silica disks. This approach allowed capacitive detection of the film's dynamics: film thickness variations change the capacitance between the plates. In previous works, we have used a few-nanometers-thick film adsorbed on a single on-chip silica microdisk allowing
 a reduction in confinement radius of three orders of magnitude together with optical detection of the film's dynamics.
\cite{Harris2016,he2019strong,McAuslan2016,Sachkou2019}.
Such disks, however, become unsuitable for confinement below $\sim \SI{10}{\micro \meter} $ micron. Using high refractive index materials such as silicon or gallium arsenide instead, the disks can be miniaturized down to $\sim \SI{1}{\micro \meter} $ radii \cite{gil-santos_high-frequency_2015}. These set-ups span the hatched region in Fig. \ref{fig:omshift_r_d_band_structure}(a); it can be seen that reaching the single-phonon nonlinear regime then requires damping rates in the millihertz regime. This is a challenging condition, yet such low damping rates---with corresponding mechanical quality factors $Q=\om / \Gamma$ in excess of $10^5$---have been experimentally observed in Ref. \cite{PhysRevB.39.2703}, albeit for larger millimeter dimensions. For bulk modes in superfluid $^4$He, quality factors in excess of $10^8$ have been measured \cite{DeLorenzo2017}. 

Further size reduction could be achieved through the use of phononic crystal cavities\cite{hybridcrystal,PhysRevLett.75.3106,thirdsoundpenrosecrystal}. When condensed on a patterned silicon substrate, the superfluid third-sound wave experiences a periodic modulation of its speed of sound and a periodic potential \cite{condat_third-sound_1985}.
This provides the possibility to engineer phononic crystal band structures that contain frequency ranges, or band gaps, for which wave propagation in the crystal is not allowed. In analogy to \afterpage{%
    \clearpage
    \thispagestyle{empty}
    \begin{landscape}
        \centering 
        \resizebox{1.4\textwidth}{!}{%
\begin{tabular}{@{}lllccccccccc@{}}
 & & & & & $\blacktriangledown$ & & & & & & \\ 
\toprule
Class & Resonator & Nonlinearity & $\alpha$ {[}\si[per-mode=fraction]{\newton \per \cubic \meter}{]} & $\xzpf$ {[}m{]} & $\frac{ \domo}{2 \pi}$ {[}\si[per-mode=fraction]{\hertz}{]} & $\frac{\om}{2 \pi}$ {[}Hz{]} & $\meff$ {[}kg{]}  & $\frac{\Gamma}{2 \pi}$ {[}Hz{]} & $\frac{ \domo }{\Gamma}$ & Year & Ref. \\ \midrule
\rowcolor{superfluidcolor}  
 Superfluid &  Third sound (R=\SI{100}{\nano \meter}, d=\SI{5}{\nano \meter}, $\radnodes =$ 10) & Intrinsic & \num{3E+16} & \num{2E-12} & \num{8E+02} & \num{4E+08} & \num{9E-21} &  &  &   &  \\
 \rowcolor{superfluidcolor}  
 Superfluid &  Third sound (R=\SI{100}{\nano \meter}, d=\SI{5}{\nano \meter}, $\radnodes =$ 1) & Intrinsic & \num{2 (1)E+16} & \num{5E-13} & \num{7 (4) E+00} & \num{5E+07} & \num{6E-19} &  &  &   &  \\
Bottom-up resonator & Graphene sheet & Intrinsic & \num{2E+16} & \num{3E-13} & \num{9E-01} & \num{2E+08} & \num{4E-19} & \num{4E+04} & \num{7E-05} & 2011 & \cite{Eichler:2011aa} \\
Bottom-up resonator & Carbon nanotube & Intrinsic & \num{6E+12} & \num{2E-12} & \num{3E-01} & \num{3E+08} & \num{8E-21} & \num{5E+03} & \num{2E-04} & 2011 & \cite{Eichler:2011aa} \\
\rowcolor{superfluidcolor}  
 Superfluid &  Third sound (R=\SI{1}{\micro \meter}, d=\SI{5}{\nano \meter}, $\radnodes =$ 10) & Intrinsic & \num{3E+18} & \num{5E-14} & \num{8E-02} & \num{4E+07} & \num{9E-17} &  &  &   &  \\
MEMS & Gold-covered d.c. Si beam & Chemical bond & \num{1E+20} & \num{7E-15} & \num{1E-03} & \num{2E+06} & \num{1E-13} & \num{5E+02} & \num{8E-06} & 2016 & \cite{Huang2016} \\
\rowcolor{superfluidcolor}  
 Superfluid &  Third sound  R=\SI{1}{\micro \meter}, d=\SI{5}{\nano \meter}, $\radnodes =$ 1)  & Intrinsic & \num{2 (1) E+18} & \num{2E-14} & \num{7 (4) E-04} & \num{5E+06} & \num{6E-15} &  &  &   &  \\
NEMS & Driven graphene sheet & Intrinsic & \num{1E+15} & \num{1E-13} & \num{5E-04} & \num{1E+07} & \num{6E-17} & \num{1E+05} & \num{1E-08} & 2017 & \cite{Davidovikj2017} \\
Levitated particle & Levitated silica nanoparticle & Optical trap & \num{-1E+07} & \num{5E-12} & \num{3E-05} & \num{1E+05} & \num{3E-18} & \num{8E-04} & \num{1E-01} & 2017 & \cite{Ricci:2017aa} \\
NEMS & Graphene-coupled $\mathrm{Si}_{3}\mathrm{N}_{4}$ membrane & Graphene & \num{8E+21} & \num{7E-16} & \num{9E-06} & \num{3E+06} & \num{6E-12} & \num{7E+02} & \num{4E-08} & 2020 & \cite{Singh2020} \\
\rowcolor{superfluidcolor}  
 Superfluid &  Third sound  (R=\SI{10}{\micro \meter}, d=\SI{5}{\nano \meter}, $\radnodes =$ 10)  & Intrinsic & \num{3E+20} & \num{2E-15} & \num{8E-06} & \num{4E+06} & \num{9E-13} &  &  &   &  \\
NEMS & Metallized SiC beam & Intrinsic & \num{5E+14} & \num{4E-14} & \num{6E-06} & \num{9E+06} & \num{6E-16} & \num{1E+03} & \num{2E-08} & 2006 & \cite{Kozinsky2006} \\
NEMS & D.c. $\mathrm{Si}_{3}\mathrm{N}_{4}$ beam & Qubit & \num{8E+15} & \num{2E-14} & \num{5E-06} & \num{6E+07} & \num{4E-16} & \num{1E+03} & \num{1E-08} & 2010 & \cite{Suh2010} \\
Levitated particle & Levitated particle & Optical trap & \num{-8E+03} & \num{2E-11} & \num{2E-06} & \num{7E+04} & \num{5E-19}  & \num{6E+02} & \num{9E-09} & 2019 & \cite{doi:10.1063/1.5116121} \\
\rowcolor{superfluidcolor}  
 Superfluid &  Third sound (R=\SI{10}{\micro \meter}, d=\SI{5}{\nano \meter}, $\radnodes =$ 1)  & Intrinsic & \num{2 (1) E+20} & \num{5E-16} & \num{7 (4) E-08} & \num{5E+05} & \num{6E-11} &  &  &   &  \\

NEMS & D.c. four-layer piezoelectric beam & Intrinsic & \num{8E+14} & \num{9E-15} & \num{3E-08} & \num{1E+07} & \num{1E-14} & \num{1E+04} & \num{7E-12} & 2013 & \cite{Matheny:2013aa} \\

NEMS & D.c. Si beam & Intrinsic & \num{5E+13} & \num{1E-14} & \num{1E-08} & \num{5E+07} & \num{8E-16} & \num{8E+03} & \num{4E-12} & 2016 & \cite{Sansa:2016aa} \\
NEMS & D.c. $\mathrm{Si}_{3}\mathrm{N}_{4}$ /Nb bilayer beam & Intrinsic & \num{2E+11} & \num{4E-14} & \num{1E-09} & \num{1E+06} & \num{7E-15} & \num{2E+01} & \num{3E-10} & 2014 & \cite{Hocke:2014aa} \\
MEMS & Suspended InP membrane & Intrinsic & \num{4E+13} & \num{9E-15} & \num{1E-09} & \num{1E+06} & \num{1E-13} & \num{2E+02} & \num{2E-11} & 2013 & \cite{Antoni:2012aa} \\
NEMS & D.c. $\mathrm{Si}_{3}\mathrm{N}_{4}$ beam & Intrinsic & \num{1E+10} & \num{3E-14} & \num{4E-11} & \num{9E+05} & \num{1E-14} & \num{2E+00} & \num{9E-11} & 2018 & \cite{Maillet:2018aa} \\
MEMS & Metallic plate Casimir resonator & Intrinsic & \num{3E+09} & \num{6E-15} & \num{1E-14} & \num{2E+03} & \num{1E-10} & \num{3E-01} & \num{2E-13} & 2001 & \cite{Chan2001} \\
AFM & Commercial van der Waals cantilever & Intrinsic & \num{2E+09} & \num{4E-15} & \num{3E-15} & \num{4E+04} & \num{1E-11} & \num{1E+03} & \num{6E-18} & 2002 & \cite{Lee2002} \\ \bottomrule
\end{tabular}%
}
        \captionof{table}{Overview of nonlinearities in various systems, sorted in order of decreasing single-phonon nonlinear frequency splitting $\domo =\frac{3 \alpha \xzpf^4}{\hbar}$. Tabulated values for third sound (highlighted) use rotationally invariant modes ($\azorder = 0$) with superfluid density $\rhosf = \SI{145}{kg/m^3} \approx \rhohe$  and the van der Waals coefficient for silica $\avdw = \SI{2.65E-24}{m^5 . s^{-2}} $ \cite{PhysRevLett.30.1122}. For the third-sound resonators, the effect of higher-order contributions from a cubic nonlinearity ($\beta \neq 0$) reduce the effective Duffing nonlinearity to respectively $\alef = 0.6 \alpha$ and $\alef = 0.98 \alpha$ for $\radnodes = 1$ and $10$, in parentheses.\\
        D.c.: doubly clamped. 
 \label{tab:overview} }
    \end{landscape}
    \clearpage
} photonic crystals and solid-state phononic crystals, we find that third sound modes can be trapped or confined in a defect in the crystal as long as its frequency lies in the band gap, as shown in Fig. \ref{fig:omshift_r_d_band_structure}(b). Thereby, phononic crystal cavities could enable confinement down to and below hundreds of nanometers (Fig. \ref{fig:omshift_r_d_band_structure}a, gray).

Here, as an example, we calculate the band structure for a \SI{11}{\nano \metre} thick superfluid film condensed on a suspended silicon slab perforated with \SI{55}{\nano \metre} diameter holes and a $\SI{100}{\nano \metre}$ lattice constant, as shown in Fig. \ref{fig:omshift_r_d_band_structure}(b). The hydrodynamic third sound equations are solved  through finite element method (FEM) simulation (see Supplementary Information and Ref. \cite{Forstner2019}), showing a complete band gap at around \SI{30}{\mega \hertz}. When a hole is removed to form a central defect, it confines a \SI{30}{\mega \hertz} acoustic mode. This trapped mode closely resembles the ($\azorder=0,\radnodes =1$) $R=\SI{56}{\nano\metre}$ circular Bessel mode (see Supplementary Information). From Eq. (\ref{eq:domofull}), the corresponding single-phonon frequency shift is $\domo[\alef]/2 \pi=\SI{35}{\hertz}$ (indicated with an X-mark in Fig. \ref{fig:omshift_r_d_band_structure}a), so that a quality factor of $\num{9 E+5}$ is required to reach the single-phonon nonlinear regime. 

Unambiguous identification of all damping mechanisms in third sound remains an open problem in the field \cite{Hoffmann:2004aa,2006PhDT229W,Bhattacharjee1982,Bergman1969,Buck1981}. The known main dissipation channels are thermal dissipation that arises due to the temperature gradient between the peaks and troughs of the sound wave, dissipation due to  interactions with pinned-vortices \cite{penanen_model_2002,Hoffmann:2004aa}, and radiation \cite{Harris2016,Sachkou2019,he2019strong} (or clamping \cite{nguyen_ultrahigh_2013}) losses.

Aside from allowing sub-micron confinement, a virtue of the phononic crystal architecture is the ability to control the radiation loss associated with imperfect reflection of the wave at the confining boundary. Our model confirms (see Supplementary Information) that the radiation damping decreases strongly  as the number of cells in the phononic crystal lattice increases. Indeed, for the $\SI{30}{\mega \hertz}$ mode in Fig. \ref{fig:omshift_r_d_band_structure} the threshold for single-phonon resolution $Q= \num{9E5}$ is reached for just $10$-cell lattices.

Thermal dissipation occurs as the result both of  evaporation and recondensation within the superfluid wave itself~\cite{Atkins1959}, and of irreversible heat flow through the film and substrate~\cite{Atkins1959, Bergman1969,Bergman1971} (similar to thermoelastic losses in micromechanical resonators~\cite{lifshitz_thermoelastic_2000}). These mechanisms were modeled for a plane wave by Atkins in Ref.~\cite{Atkins1959}. Using this model (see Supplementary Information) we find that these mechanisms are greatly suppressed at low temperatures (where the temperature gradients in the wave are reduced) and also for high mechanical frequencies. The suppression with increasing mechanical frequency is perhaps surprising. However, it can be understood since if the frequency is significantly higher than the rate of thermal equilibration, then heat has insufficient time to flow between troughs and peaks. A similar phenomenon has been observed for thermoelastic losses~\cite{schmid2016fundamentals}. Combined, we predict that thermal losses will not be a significant source of dissipation for strongly confined films at low temperatures. For the specific example of the phononic-crystal structure modeled in Fig.~\ref{fig:omshift_r_d_band_structure}(b), we find that the thermal dissipation rate is suppressed beneath the  frequency shift due to a single-phonon even at temperatures as high as 0.5~K. Indeed, sub-hundred millikelvin thermalisation is not uncommon for third sound~\cite{Baierlein1997,PhysRevB.39.2703,Ellis1993,Ellis2004,Hoffmann:2004aa,Kono1981,PhysRevLett.81.152,PhysRevLett.32.147,Schechter:1998aa}, in which case Atkins' model predicts that the thermal dissipation dominated superfluid quality factor $Q_{\mathrm{thermal}}$ would exceed $10^{10}$.

Dissipation resulting from interactions with quantized vortices pinned to the resonator surface is thought to arise due to both vortex-normal fluid interactions and vortex dimple drag \cite{penanen_model_2002,Hoffmann:2004aa}.
These damping mechanisms require large pinned vortex densities in order to account for the observed dissipation, with densities on the order of $10^{13}$~cm$^{-2}$ estimated in Ref. \cite{penanen_model_2002}. Recent experimental work, however, shows that coherent vortex-vortex interactions dominate pinning when the superfluid film is confined at hundred micron-scales on smooth microfabricated resonators~\cite{Sachkou2019}. 
The vortex distribution then evolves towards its lowest energy, vortex-free, configuration over minute periods. As the resonator radius is scaled down, it is expected that the influence of pinning sites will be further reduced since the sound-vortex coupling scales inversely with the resonator area~\cite{Forstner2019}, causing vortices to be dislodged from pinning sites and their subsequent annihilation.
 
Liquid helium only remains superfluid as long as the fluid particle velocity is less than the superfluid critical velocity. It is interesting then to ask whether there are any fundamental limits to how strongly the superfluid film may be confined, both in thickness and radius, before superfluidity breaks down, and whether these constrain the possibility of entering the single phonon nonlinear regime. In the Supplementary Information we consider two cases, first a third sound mode cooled to its motional ground state, and second a third sound mode in thermal equilibrium with its environment. For both cases, we find that a 10~nm thick film would need to be confined to a radius beneath its thickness for the fluid particle velocity to exceed the critical velocity (a parameter regime outside the validity of our model). We conclude, therefore, that the breakdown of superfluidity places no constraints on reaching the single phonon strong coupling regime with the hundred-nanometer-and-above diameter superfluid resonators considered here.
 
\section*{DISCUSSION}\label{sec:discussion}
We have shown that third-sound resonances (surface oscillations of two-dimensional superfluid helium) are intrinsically strongly nonlinear, and have identified the specific parameters that allow maximization of the nonlinearities: film thickness, confinement radius, and radial and rotational mode symmetry. We showed that the cubic nonlinearity can be treated analytically as an effective reduction of the quartic nonlinearity, which diminishes rapidly for surface waves with an increasing number of radial nodes.

We calculated the expected output spectrum in the presence of decoherence and predict that single-phonon nonlinear frequency shifts exceeding even those of graphene sheets and carbon nanotubes by orders of magnitude may be possible. Combined with the intrinsically low dissipation of motional states in superfluid \cite{PhysRevB.39.2703}, and phononic crystal cavities providing sub-100nm confinement and strongly  suppressed radiation damping, this may well open the door to the single-phonon nonlinear regime where a single phonon can shift the resonance frequency by more than the mechanical linewidth. 

Our results dovetail recent theoretical proposals that lay out how strong Duffing nonlinearities can be used for quantum control and metrology \cite{Rips2012,Lu2015,Babourina-Brooks2008,Buks2006,Momeni2019,Woolley2008}. Ultimately, they provide a pathway towards new tests of quantum macroscopicity and new tools for quantum technologies, where mechanical resonators can function not only as oscillators, memories, and interfaces, but also as qubits.

\section*{ACKNOWLEDGMENTS}
This work was funded by the US Army Research Office through grant number W911NF17-1-0310 and the Australian Research Council Centre of Excellence for Engineered Quantum Systems (EQUS, project number CE170100009). W.P.B. and C.G.B respectively acknowledge Australian Research Council Fellowships FT140100650 and DE190100318. L.T. is supported by the National Science Foundation (USA) under Award No. 1720501 and No. 2006076.

\section*{REFERENCES}


\bibliographystyle{ieeetr}
\renewcommand\refname{}

\bibliography{nonlinsfbibliography_main}{}
\end{document}


\twocolumn[
\vspace{-2cm}
\maketitle 
]

\section{Nonlinear spring constants for a superfluid thin film}

As detailed in the main text, the linear spring constant $k$, cubic nonlinearity $\beta$ and quartic (Duffing) nonlinearity $\alpha$ for a superfluid surface wave of amplitude $\amp\left[ r, \theta \right]$ are given by
\begin{equation} \label{eq:k1}
k=\frac{3 \,\rhosf\, \avdw }{d^4} \int_{0}^{2\pi}  \int_0^R \, \frac{\amp^2\left[ r, \theta \right]}{\amp^2\left[ R, 0\right]} \, r \,\mathrm{d}r\, \mathrm{d}\theta ,
\end{equation}
\begin{equation} \label{eq:beta1}
\beta=-\frac{6 \,\rhosf\, \avdw }{d^5} \int_{0}^{2\pi}   \int_0^R \, \frac{\amp^3\left[ r, \theta \right]}{\amp^3\left[ R, 0\right]}\, r \,\mathrm{d}r\, \mathrm{d}\theta
\end{equation}
and
\begin{equation} \label{eq:alpha1}
\alpha=\frac{10 \,\rhosf\, \avdw }{d^6} \int_{0}^{2\pi}  \int_0^R \, \frac{\amp^4\left[ r, \theta \right]}{\amp^4\left[ R, 0\right]}\, r \,\mathrm{d}r\, \mathrm{d}\theta .
\end{equation}
In order to reveal the explicit dependence of the (non)linear spring constants on $R$, $d$, $\azorder$ and $\radnodes$, we evaluate the integrals further.

The integrals in equations (\ref{eq:k1}--\ref{eq:alpha1}) can be written jointly as a function $\Cint$ with $p=2,3$ and $4$, with
\begin{equation} \label{eq:intoverlap}
\Cint:= \int_{0}^{2\pi}  \int_0^R \, \frac{J_{\azorder}^p \left[\zeta_{\azorder , \radnodes} \frac{r}{R}\right]\cos ^p \left( \azorder  \theta\right)}{J_{\azorder}^p\left[\zeta_{\azorder , \radnodes}\right]}\, r \,\mathrm{d}r\, \mathrm{d}\theta .
\end{equation}
The integral over the angular coordinate $\theta$ in Eq. (\ref{eq:intoverlap}) is
\begin{align} \label{eq:intangular}
& \int_{0}^{2\pi} \cos ^p \left( \azorder  \theta\right) \, \mathrm{d} \theta  \nonumber  \\ & = 2 \pi \Bigg(
\delta_{\azorder 0} + (1- \delta_{\azorder 0}) (1- \delta_{p 3}) \frac{(p-1)!!}{p!!} \Bigg) \\ & = \left\{ \pi ( 1+ \delta_{\azorder 0} ), \, 2 \pi \delta_{\azorder 0},  \, \pi \frac{3+5 \delta_{\azorder 0} }{4} \right\} \, \, \text{for } \, p= \{ 2,3,4\} ,
\end{align}
where we have introduced the Kronecker delta function $\delta$.
Observe here: the reduction of this integral to $2 \pi \delta_{\azorder 0}$ for $p=3$ implies that $\Phi_{ \azorder  \neq 0,\radnodes }^{(3)} =0$, so the cubic nonlinearity $\beta$ vanishes for all but the zeroth-order ($ \azorder =0$) superfluid modes.

We can rewrite the remainder of Eq. (\ref{eq:intoverlap}) by substitution of the integrand:
\begin{align} 
& J_{\azorder}^{-p}\left[\zeta_{\azorder , \radnodes}\right] \int_0^R \, J_{\azorder}^p \left[\zeta_{\azorder , \radnodes} \frac{r}{R}\right]  \, r \, \mathrm{d}r \nonumber \\
& =  \frac{ \int_0^{\zeta_{\azorder , \radnodes}}  J_{\azorder}^{p}\left[q\right] \, q \, \mathrm{d}q}{\zeta_{\azorder , \radnodes}^{2}  J_{\azorder}^{p}\left[\zeta_{\azorder , \radnodes}\right] } \, R^2 \nonumber \\
& :=  \cint \, R^2 \, .
\end{align}
It follows immediately that all spring constants $k, \beta$ and $\alpha$ scale with the square of the confinement radius:
\begin{align} 
\Cint &= 2 \pi \left(
\delta_{\azorder 0} + (1- \delta_{\azorder 0}) (1- \delta_{p 3}) \frac{(p-1)!!}{p!!} \right) \cint \, R^2 \label{eq:intfullsubst} .
\end{align}
The constants $\cint =  \frac{ \int_0^{\zeta_{\azorder , \radnodes}}  J_{\azorder}^{p}\left[q\right] \, q \, \mathrm{d}q}{\zeta_{\azorder , \radnodes}^{2}  J_{\azorder}^{p}\left[\zeta_{\azorder , \radnodes}\right] } $ are tabulated in Table \ref{tab:int} for the three lowest mode orders $ \azorder $ and $\radnodes $.	

It is worth noting here that for $p=2$, i.e., for the spring constant $k$, a closed-form expression exists for the integral $ \int_0^{\zeta_{\azorder , \radnodes}}  J_{\azorder}^{p}\left[q\right] \, q \, \mathrm{d}q $. In that case, we find
\begin{align} 
 \int_0^{\zeta_{\azorder , \radnodes}}  J_{\azorder}^{2}\left[q\right] \, q \, \mathrm{d}q & = \frac{\zeta_{\azorder , \radnodes}^2}{2}  \left(  J_{ \azorder -1} ^{2}\left[\zeta_{\azorder , \radnodes}\right] +  J_{\azorder}^{2}\left[\zeta_{\azorder , \radnodes}\right] \right) \nonumber \\ & -  \azorder  \, \zeta_{\azorder , \radnodes}  J_{ \azorder -1} \left[\zeta_{\azorder , \radnodes}\right]   J_{\azorder}\left[\zeta_{\azorder , \radnodes}\right] \nonumber \\ &= \frac{ \zeta_{\azorder , \radnodes}^2 -  \azorder ^2}{2}  J_{ \azorder } ^{2}\left[\zeta_{\azorder , \radnodes}\right] \nonumber ,
\end{align}
where we took advantage of the Bessel function recurrence relations and our definition $J_{\azorder}^{'}[\zeta_{\azorder , \radnodes}]=0$ so that $J_{\azorder - 1}[\zeta_{\azorder , \radnodes}]=\frac{\azorder}{\zeta_{\azorder , \radnodes}} J_{\azorder}[\zeta_{\azorder , \radnodes}]$.
Then,
\begin{align} \label{eq:intp2}
\phi^{(2)}_{\azorder , \radnodes} =  \half \left( 1- \frac{ \azorder ^2}{\zeta_{\azorder , \radnodes}^2} \right)  .
\end{align}
Since $\zeta_{\azorder , \radnodes}$ is always larger than $ \azorder $, we have
\begin{align} \label{eq:phi2scale}
0 < \phi^{(2)}_{\azorder , \radnodes} \leq \half 
\end{align}
and up to first order  $\phi^{(2)}_{\azorder , \radnodes} $ is independent of $\azorder$ and $\radnodes$. While no closed form exists for $\phi^{(3)}_{\azorder , \radnodes} $, it too, is found to be bounded:
\begin{align} \label{eq:phi3scale}
0< \lvert \phi^{(3)}_{\azorder , \radnodes} \rvert \leq \lvert \phi^{(3)}_{0 , 1} \rvert = 0.44 .
\end{align}
The function $\phi^{(4)}_{\azorder , \radnodes} $ does not converge for $\radnodes \to \infty $, but it grows sufficiently slowly that for the first twenty mode numbers it is contained in a relatively small interval:
\begin{align} \label{eq:phi4scale}
0< \phi^{(4)}_{\azorder , \radnodes} \leq \phi^{(4)}_{0 , 20} = 2.3 \, \,  \, \, \, \, \, \, \, \, (\azorder , \radnodes \leq 20) \, .
\end{align}
The observations (\ref{eq:phi2scale}--\ref{eq:phi4scale}) are important, because the (non)linear spring constants depend on the mode numbers $\azorder$ and $\radnodes$ through the function $\phi^{(p)}_{\azorder , \radnodes}$.

\begin{table}[h]
\centering
\begin{tabular}{@{}lrrr@{}}
\toprule
\multicolumn{4}{c}{${\phi}^{(2)}_{\azorder , \radnodes}$} \\ \midrule \midrule
 & \textit{$\radnodes =$ 1} & \textit{$\radnodes =$ 2} & \textit{$\radnodes =$ 3} \\
\textit{ $\azorder$ =0} & $ 1/2 $ & $1/2$ & $1/2$ \\
\textit{ $\azorder$ =1} & \num{0.353	} & \num{0.482	} & \num{0.493} \\
\textit{ $\azorder$ =2} & \num{0.286	} & \num{0.456	} & \num{0.480} \\  \midrule
\multicolumn{4}{c}{${\phi}^{(3)}_{\azorder , \radnodes}$} \\ \midrule \midrule
\textit{ $\azorder$ =0} & \num{- 0.437	} & \num{0.259	} & \num{- 0.236} \\
\midrule
\multicolumn{4}{c}{${\phi}^{(4)}_{\azorder , \radnodes}$} \\ \midrule \midrule
\textit{ $\azorder$ =0} & \num{1.28	 } & \num{1.48	} & \num{1.61} \\
\textit{ $\azorder$ =1} & \num{0.290	} & \num{0.837	} & \num{1.03} \\
\textit{ $\azorder$ =2} & \num{0.223	} & \num{0.704	} & \num{0.891} \\ \bottomrule
\end{tabular}
\caption{Coefficients ${\phi}^{(p)}_{\azorder , \radnodes} =  \frac{ \int_0^{\zeta_{\azorder , \radnodes}}  J_{\azorder}^{p}\left[q\right] \, q \, \mathrm{d}q}{\zeta_{\azorder , \radnodes}^{2}  J_{\azorder}^{p}\left[\zeta_{\azorder , \radnodes}\right] }$ for $p= \{ 2,3,4\}$ with $\zeta_{\azorder , \radnodes}$ the $\radnodes^{\text{th}} $ zero of  $J_{\azorder}^{'}$. 
}
\label{tab:int}
\end{table}

With Eq. (\ref{eq:intoverlap}), (\ref{eq:intfullsubst}) and (\ref{eq:intp2}), we can then expose the dependence of the (non)linear spring constants on the film thickness $d$ and confinement radius $R$:

\begin{align} \label{eq:k}
k &= \left( 1+ \delta_{\azorder 0} \right) 3 \pi \,\rhosf \avdw  \, {\phi}^{(2)}_{\azorder , \radnodes} \, \frac{R^2}{d^4} \\
& = \left( 1+ \delta_{\azorder 0} \right) \frac{3 \pi}{2}  \rhosf \avdw \left( 1- \frac{ \azorder ^2}{\zeta_{\azorder , \radnodes}^2} \right) \, \frac{R^2}{d^4} \, ,
\end{align}

\begin{align} \label{eq:beta}
\beta = - \delta_{\azorder 0} \, 12 \pi \,  \rhosf \avdw   \, {\phi}^{(3)}_{0 , \radnodes} \, \frac{R^2}{d^5} 
\end{align}
and
\begin{align} \label{eq:alpha}
\alpha = \left( 3+ 5 \delta_{\azorder 0} \right) \frac{5 \pi}{2} \,\rhosf \avdw  \, {\phi}^{(4)}_{\azorder , \radnodes} \, \frac{R^2}{d^6} \, .
\end{align}

\section{Single-phonon transition resonances from perturbation theory}\label{sec:effective}
The nonlinear Hamiltonian can be written, with $\lambda \ll 1$ a dimensionless parameter, as the sum $H = H_0+\lambda H_1$ of the unperturbed Hamiltonian
\begin{equation}
H_0 = \frac{p^2}{2 \meff} + \frac{1}{2}k\,x^2
\end{equation}
and the perturbation
\begin{equation}
H_1 = \frac{1}{3} \beta x^3 + \frac{1}{4} \alpha x^4 .
\end{equation}
 Then, within first order perturbation theory,
\begin{align}
\lambda E_n^{(1)}&=\left\langle n\right|\lambda H_1\left|n\right\rangle  \nonumber \\
&= \lambda \frac{\beta}{3} \xzpf^3 \left\langle n\right|(a+\ad)^3\left|n\right\rangle + \lambda^2 \frac{\alpha}{4} \xzpf^4 \left\langle n\right|(a+\ad)^4\left|n\right\rangle \nonumber \\
&=0+\lambda \frac{3 \xzpf^4 \alpha}{2} (n^2+n+\half)
\end{align}
and through second order perturbation theory,

\begin{align}
\lambda^2 E_n^{(2)}&= \sum_{k \neq n} \frac{|\left\langle k\right|\lambda H_1\left|n\right\rangle|^2}{E_n^{(0)}-E_k^{(0)}} \nonumber \\
&=\frac{\lambda^2}{\hbar \om} \left( -\frac{\alpha^2}{8}\xzpf^8 \left( 34 n^3+51n^2+59n+21  \right) \right. \nonumber \\
&- \left. \frac{\beta^2}{9}\xzpf^6 \left( 30 n^2+30n+11  \right) 
\right) .
\end{align}
The energy eigenvalues of the Fock states of the harmonic oscillator then become 

\begin{align}
E_n &= E_n^{(0)} + \lambda E_n^{(1)} + \lambda^2 E_n^{(2)} \nonumber  \\ \nonumber
&= \hbar \om (n+\frac{1}{2}) \\ \nonumber
&+ \lambda \frac{3 \xzpf^4 \alpha}{2} \left( n^2+n+\half \right) \\ \nonumber
&- \lambda^2 \frac{\xzpf^6 \beta^2}{9 \hbar \om} \left( 30 n^2+ 30n+11 \right) \\ 
&- \lambda^2 \frac{\xzpf^8 \alpha^2}{8 \hbar \om} \left( 34 n^3+51n^2+59n+21 \right) \label{eq:hamanalyt1} 
\end{align}
with transition energies
\begin{align}
E_{n+1}-E_n &= \om \hbar + \lambda 3 (n+1) \xzpf^4 \left( \alpha - \lambda \frac{20}{9} \frac{\xzpf^2 \beta^2}{\hbar \om}\right) \nonumber \\
&- \lambda^2 \frac{3 \xzpf^8}{4 \hbar \om} (24+17n (n+2)) \alpha^2.
\end{align}

For the systems considered here, we can identify the small factors $\frac{\alpha \xzpf^4}{\om \hbar}\sim\frac{\alpha \xzpf^2}{k}$ and $\frac{\beta \xzpf^3}{\om \hbar}\sim\frac{\beta \xzpf}{k}$. Indeed, they are factors of increasingly high order appearing in the Taylor expansion of the Van der Waals potential in main text Eq. (6) and (8): noting that $ \alpha\, x^3\ll\beta\, x^2\ll k\,x $, we necessarily have $\frac{\alpha \xzpf^4}{\om \hbar} \ll \frac{\beta \xzpf^3}{\om \hbar}$.

For example, for the mode discussed in Fig. 6 of the main text and in SI Fig. 2 and 3, $\frac{\beta \xzpf^3}{k \xzpf^2} = \num{5E-4}$ and$\frac{\alpha \xzpf^4}{k \xzpf^2} = \num{8E-7}.$

We can thus approximate 

\begin{align} \label{eq:deltaEn}
\Omega[n] &= \frac{E_{n+1} - E_{n}}{\hbar} \nonumber \\
&= \om + \frac{\lambda 3 (n+1) \xzpf^4}{\hbar} \left( \alpha - \lambda \frac{20}{9} \frac{\xzpf^2 \beta^2}{\hbar \om}  \right) \nonumber ,
\end{align}
which allows us to write the cubic nonlinearity as an effective modification to the Duffing nonlinearity.

 \section{Spectral function calculation with Lindblad master equation}\label{sec:solution}
We numerically solve the spectral function of the nonlinear resonator, using the full Lindblad master equation of the open quantum system comprising the resonator and its environment. 

The Hamiltonian of the nonlinear mechanical mode $H = \frac{p^2}{2 \meff} + \frac{1}{2}k\,x^2+\frac{1}{3} \beta x^3 + \frac{1}{4} \alpha x^4 $ can be written in terms of the eigenstates
$\left|j\right\rangle $ and eigenvalues $E_{j}$ as $H =\sum_{j} E_{j}\left|j\left\rangle \right\langle j\right|$. The eigenstates and eigenvalues are obtained by numerically discretizing and diagonalizing the Hamiltonian. 

The oscillation amplitude can be written in terms of the eigenbasis as
\begin{align}
x=\sum_{k>j} \left( x_{jk}\left|j\right\rangle \left\langle k\right|+ \, \text{h.c.} \right) +\sum_{j}x_{jj}\left|j\right\rangle \left\langle j\right| ,
\end{align}
with matrix elements $x_{jk}=\left\langle j\left|x\right|k\right\rangle $. The quantum master equation of the mechanical mode coupled to a bath of environmental modes can be derived using the standard perturbation theory approach in the eigenbasis  \cite{Gardiner2004}. Omitting the fast rotating terms in the system-bath coupling, we obtain the Lindblad master equation:
\begin{equation}
\frac{d\dens}{dt}=-i\left[H ,\dens\right]+{\cal L}\dens\label{eq:drho}
\end{equation}
with
\begin{equation}
\begin{split}
{\cal L} =\frac{\Gamma}{2} \sum_{k>j}\left|x_{jk}\right|^{2}\Big( \left( n_{{\rm th}}\left[\delta E_{kj}\right]+1\right){\cal D}_{jk}
\Big. 
\\
+ \,
\Big. 
 n_{{\rm th}}\left[\delta E_{kj}\right]{\cal D}_{kj} \Big) ,
\label{eq:lind}
\end{split}
\end{equation}
where $\dens$ is the density matrix of the mechanical mode, $n_{{\rm th}}\left[\delta E_{kj}\right]=\left(e^{\hbar\delta E_{kj}/k_{B}T}-1\right)^{-1}$
is the thermal phonon occupation number at temperature $T$ for the frequency difference $\delta E_{kj}= E_{k}- E_{j}$ between the states $k$ and $j$, and 
\begin{align}
{\cal D}_{jk} \, \dens =2\left|j\right\rangle \left\langle k\right|\dens\left|k\right\rangle \left\langle j\right|-\left|k\right\rangle \left\langle k\right|\dens-\dens\left|k\right\rangle \left\langle k\right|
\end{align}
is the Lindblad superoperator for the jump operation $\left|j\right\rangle \left\langle k\right|$.
The difference between Eq. (\ref{eq:drho}) and the standard master equation for a quantum harmonic oscillator stems from the nonlinearity in the Hamiltonian, which perturbs the equal energy level spacing.

Defining amplitude operators $\epsilon^{+}=\sum_{k>j}x_{jk}\left|j\right\rangle \left\langle k\right|$
and $\epsilon^{-}=\left(\epsilon^{+}\right)^{\dag}$, one can calculate the correlation function $G[\tau]$ of the mechanical amplitude as a function of the separation time $\tau$:
\begin{equation}
G^{}[\tau]=\left\langle \epsilon^{-}\left[t+\tau\right]\epsilon^{+}\left[t\right]\right\rangle _{t\rightarrow\infty} . \label{eq:G1tau}
\end{equation}
Here we used the Heisenberg representation for the time-dependent operator $\epsilon$ such that $\epsilon[t]=e^{i\widetilde{H}t} \, \epsilon \, e^{-i\widetilde{H}t}$ with $\widetilde{H}$ the total Hamiltonian of the mechanical mode coupled to the bath modes.

Applying the quantum regression theorem \cite{PhysRev.129.2342},  we write the correlation in Eq. (\ref{eq:G1tau}) as a trace over the Hilbert space of the mechanical mode. In doing so, we introduce the stationary density matrix of the master equation (\ref{eq:drho}) $\dens_{ss}$, which is obtained by setting $\frac{\mathrm{d} \dens}{\mathrm{d} t}=0$ in Eq. (\ref{eq:lind}). The correlation function then becomes:
\begin{align}
G^{}[\tau]={\rm Tr}_{s}\left[\epsilon^{-}e^{{\cal L}\tau}\left(\epsilon^{+}\dens_{ss}\right)\right].
\end{align}
These equations are implemented numerically \cite{Tan_1999} to find the solution of $G^{}[\tau]$. The spectral function $S_{xx}[\Omega]$ can then be obtained as the Fourier transformation
of $G^{}[\tau]$: 
\begin{align}
S_{xx}[\Omega]=\frac{1}{2\pi}\int_{-\infty}^{\infty} e^{-i\Omega\tau}G^{}[\tau] \, \mathrm{d}\tau .
\end{align}

\section{Dissipation of energy}
Energy contained in superfluid $^4$He third sound may decay through several different channels. Understanding ways energy can dissipate is especially important in the third sound resonator, since the resolution of intrinsic single-phonon statistics requires the damping $\Gamma$ to be smaller than the the single-phonon resonance shift $\domo$. Experimentally, it has been observed that $Q\cdot f$ products (the product of a resonator's quality factor and frequency) in nanomechanical systems tend to lie below $10^{16}$ \SI{}{\hertz} \cite{RevModPhys.86.1391}, although much larger $Q \cdot f$ values exceeding $10^{19}$ have been reached recently by use of phononic crystal cavities \cite{Ren2020}. It nonetheless is reasonable to require $Q f$ products for single-phonon nonlinear third-sound resonators to lie below that limit, within the regime of operation of the majority of nanomechanical resonators. 

Firstly, dissipation mechanisms involving the role of large densities of pinned vortices have been proposed to account for the observed dissipation in experiments~\cite{penanen_model_2002,Hoffmann:2004aa}. Our recent work on optomechanical detection of vortices however appears to rule out such large remnant vortex densities on smooth microfabricated third sound resonators~\cite{Sachkou2019}. In addition, the sound-vortex coupling is known from theory to increase with decreasing resonator area \cite{Forstner2019}, resulting in a further reduction of vortex pinning and therefore further reduction of vortex-induced dissipation. (See main article text.)

Secondly, in the two-fluid model of superfluid third sound, the superfluid component oscillates while the normal fluid component remains stationary. This leads to temperature gradients between the colder wave peaks and warmer troughs~\cite{Atkins1959}, and  these temperature gradients can lead to thermal energy dissipation through evaporation and recondensation of helium atoms between the peaks and troughs of the wave, as well as irreversible heat flow through the substrate~\cite{Atkins1959, Bergman1969,Bergman1971}---forming in essence an analog of the thermo-elastic damping encountered in mechanical resonators~\cite{lifshitz_thermoelastic_2000}.

Finally, a fundamental loss mechanism remains the acoustic energy radiated out of the resonator due to imperfect wave reflection at the resonator boundary---corresponding for instance to the acoustic energy lost through the pedestal in the case of a superfluid-coated  microdisk or microtoroid geometry~\cite{Harris2016,Sachkou2019,he2019strong}. This dissipation mechanisms plays the role of clamping losses in micromechanical resonators \cite{nguyen_ultrahigh_2013}, and, as in the case of solid microresonators, may be suppressed through the use of a phononic bandgap structure~\cite{tsaturyan_ultracoherent_2017,condat_third-sound_1985}. In particular, the damping $\Gamma$ decreases exponentially for modes trapped in phononic lattices with an increasing number of cells. 

We study in detail the process of thermal dissipation and acoustic radiation in the following two Sections.

\section{Thermal dissipation}

\begin{figure}[h]
\centering
\includegraphics[width=0.45\textwidth]{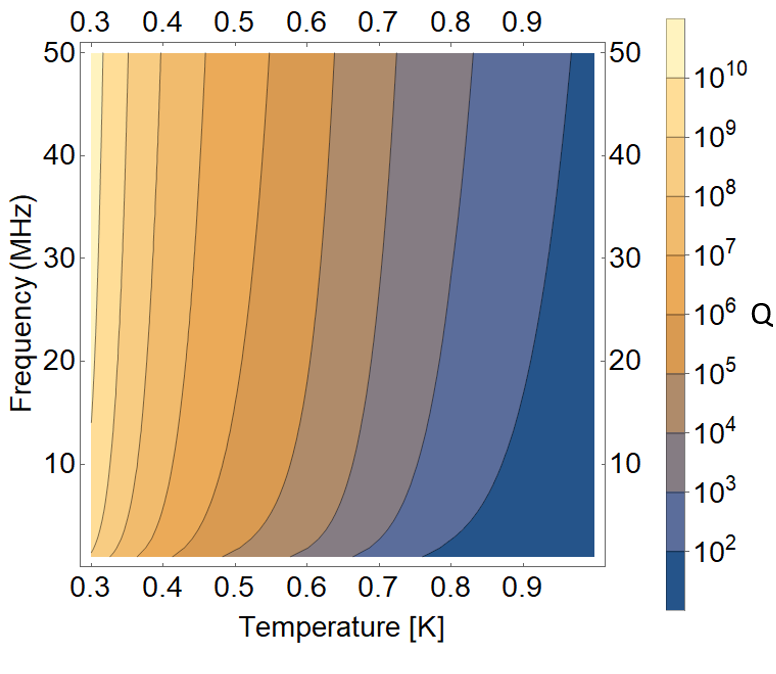}
 \caption{Third sound wave quality factor as a function of frequency and temperature for a film thickness of 11~nm. The physical properties of the film and substrate have been taken from \cite{Baker2016, Donnelly1998, Brookhaven1980, Molinari2016, Zeller1971}.}\label{fig:qcontour}
\end{figure}

In the usual limit, third sound waves are treated as an oscillation in the height of the superfluid component of the film, with the normal component viscously clamped and therefore of constant height. This approximation is valid so long as the normal-fluid penetration depth $d_p = (2 {\text v}_n/\Omega \rho_n)^{1/2}$ is larger than the film thickness $d$, where ${\text v}_n$ is the viscosity of the normal fluid and $\Omega$ is the frequency of the wave. In this case,
the motion of the sound wave creates oscillating regions of high and low superfluid-to-normal fluid ratio. This changing ratio corresponds to a change in the temperature in the thin film between crests and troughs of the third sound wave. Heat will then flow from troughs to crests, dissipating energy.

Bergman derived a solution for the complex speed of third sound plane waves $c_3$, including both evaporation-condensation and thermal damping through the substrate~\cite{Bergman1971}. 
However, Bergman's equations are highly complex and difficult to solve. In earlier work, Atkins presented an alternative simpler solution~\cite{Atkins1959}, making a series of approximations about the film thickness, thermal dissipation mechanisms and third sound frequency. In the thick film/high frequency limit (as would be the case here), attenuation of third sound is dominated by the phenomenon of evaporation and condensation of
helium atoms between the film and the gas~\cite{Bergman1971}. In this limit, the attenuation derived by Atkins differs from Bergman's more complete analysis only by a factor of 16/9~~\cite{Bergman1971}, and provides therefore a reasonable estimate of the expected magnitude of the thermal damping. Atkins' expression for the complex third sound speed $c_3$ is given by~\cite{Atkins1959}:
\begin{align}\label{eq:atkins1958}
    c_3^2 &= \frac{\frac{\rho f d}{\rho_{\rm He}} + \frac{\rho ST}{\rho_{\rm He}}\left[\left( S- \frac{\beta}{\rho_{\rm He}}\right)-i\frac{K f}{\rho_{\rm He} \Omega}\right]}{ C-i\frac{KL}{\rho_{\rm He} \Omega d}}
\end{align}
Where $\rho_{\rm He}$ and $\rho$ are the total fluid density and superfluid component density, respectively,  $d$ is the depth of the film and $f$ is the van der Waals force per unit mass at the surface of the film \cite{Baker2016}. $C$,  $S$ and $L$ are the specific heat, entropy and latent heat of evaporation of the film \cite{Donnelly1998}. $\beta$ is the slope of the vapour pressure curve and $K$ is the mass flow from evaporation both taken from the Vapour Pressure data published by Donnelly and Barenghi \cite{Donnelly1998}. Solving this equation for $c_3$ gives the quality factor $Q$ of the third sound wave via $Q=\Re(c_3)/2\Im(c_3)$.

We use Eq.~(\ref{eq:atkins1958}) here to determine the thermal dissipation dominated quality factor for a third sound wave on a silica substrate as a function of temperatures, film thicknesses and third sound frequencies.

Fig.~\ref{fig:qcontour} shows the thermal dissipation dominated $Q$ for an 11~nm thick film as a function of temperature and oscillation frequency. We chose a frequency range of 1~to~50~MHz , consistent with the range of third sound mechanical resonance frequencies expected for confinement length scales ranging from tens of nanometres to a micron with this film thickness. We choose a temperature range of 0.3 to 1.0~K, easily achievable using a standard dilution refrigerator or helium-3 cryostat. From the figure we can see that the quality factor improves dramatically with decreasing temperature, and that it increases with increasing frequency. This rise with frequency is expected from Eq.~(\ref{eq:atkins1958}), where high frequencies suppresses the complex component responsible for the damping. This effect of high frequency reducing damping also arises in thermoelastic damping where there is a characteristic time for the temperature fluctuation from a deflected beam to diffuse. At high frequency the system does not have sufficient time to respond to the change in temperature, reducing the diffusion of heat through the system and increasing the Q \cite{Schmid2016}. The thermal-dissipation-dominated quality factor exceeds $10^6$ for the entire frequency range for temperatures beneath 0.4~K.

\begin{figure}[h]
\centering
\includegraphics[width=0.45\textwidth]{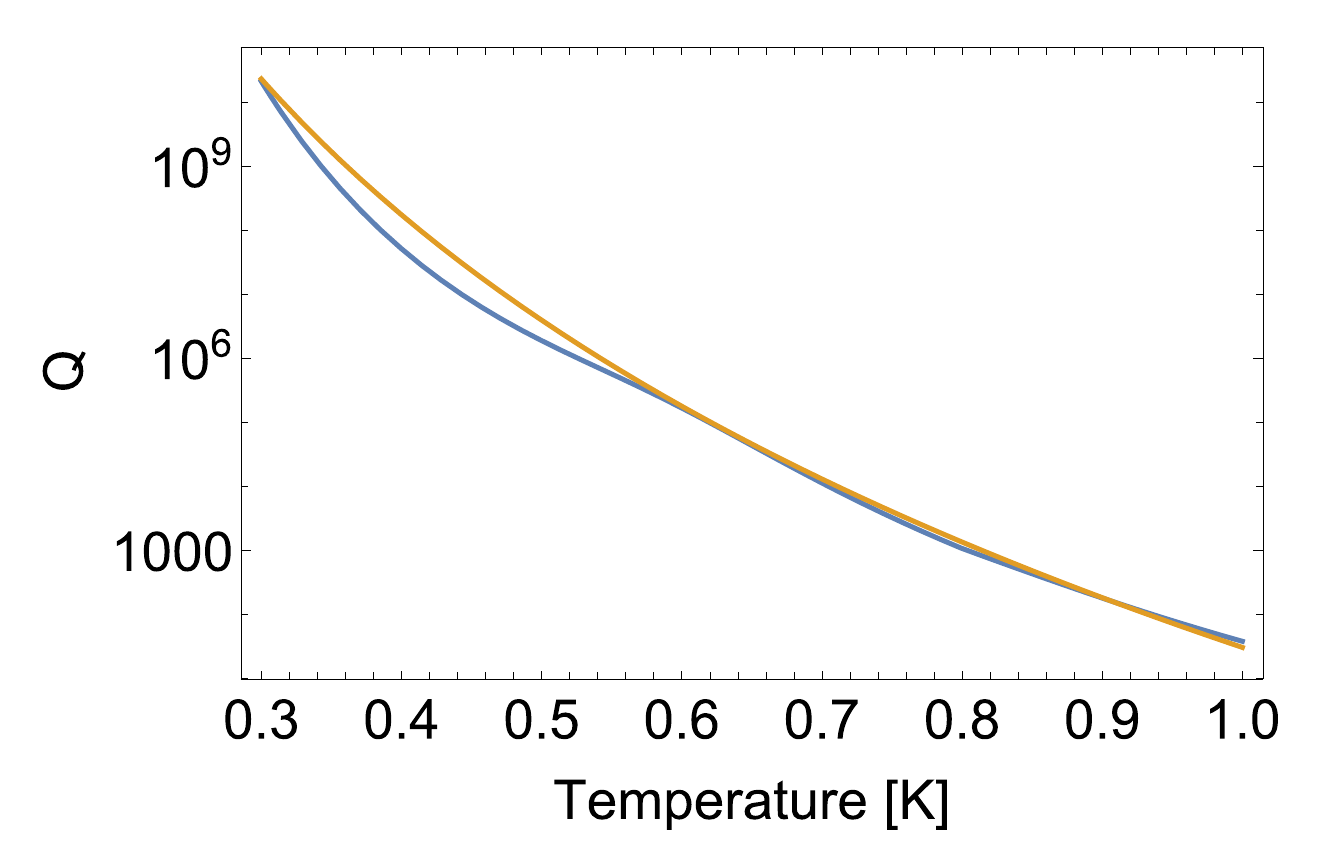}
 \caption{Third sound wave quality factor as a function of temperature at 30 MHz for a 11 nm thick film. The yellow line: rough fit to a $T^{17}$ dependence.}\label{fig:QofT}
\end{figure}

At low temperature the film has almost no normal fluid component. The difference in the proportion of superfluid component at the peaks and troughs of the third sound wave is reduced as a result, which reduces the temperature gradient. This causes the thermal damping to drop significantly as the normal fluid fraction drops. This dependence can be seen in Fig.~\ref{fig:QofT}. The third sound wave frequency and film thickness in this case are chosen to match the phononic crystal cavity mode shown in Fig.~6(b) of the main text. As can be seen, the predicted thermal-dissipation-dominated quality factor exceeds $10^6$ at temperatures beneath 0.5~K, so that (as discussed in the main text) thermal dissipation should not preclude reaching the single phonon nonlinear regime. The fit in Fig.~\ref{fig:QofT} shows a rough temperature dependence of $Q \propto T^{17}$, showing the dramatic suppression of thermal dissipation with decreasing temperature.

\begin{figure}[h]
\centering
\includegraphics[width=0.45\textwidth]{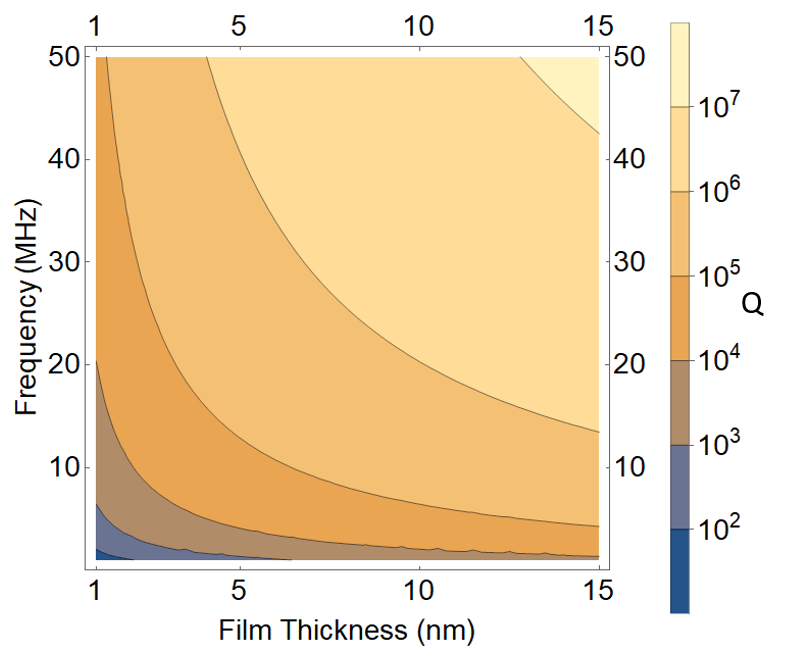}
 \caption{Third sound quality factor as a function of film thickness and frequency, at a temperature of 400 mK.}\label{fig:q_d_f}
\end{figure}

Figure~\ref{fig:q_d_f} shows the dependence of the thermal-dissipation-dominated third sound quality factor on frequency and film thickness for a fixed temperature of 0.4~K. This shows that the quality factor is predicted to decrease as the film thickness reduces. It also illustrates, again, the strong increase in quality factor predicted with increasing frequency.

\section{Superfluid phononic crystal}
\begin{figure}[h]
\centering
\includegraphics[width=0.45\textwidth]{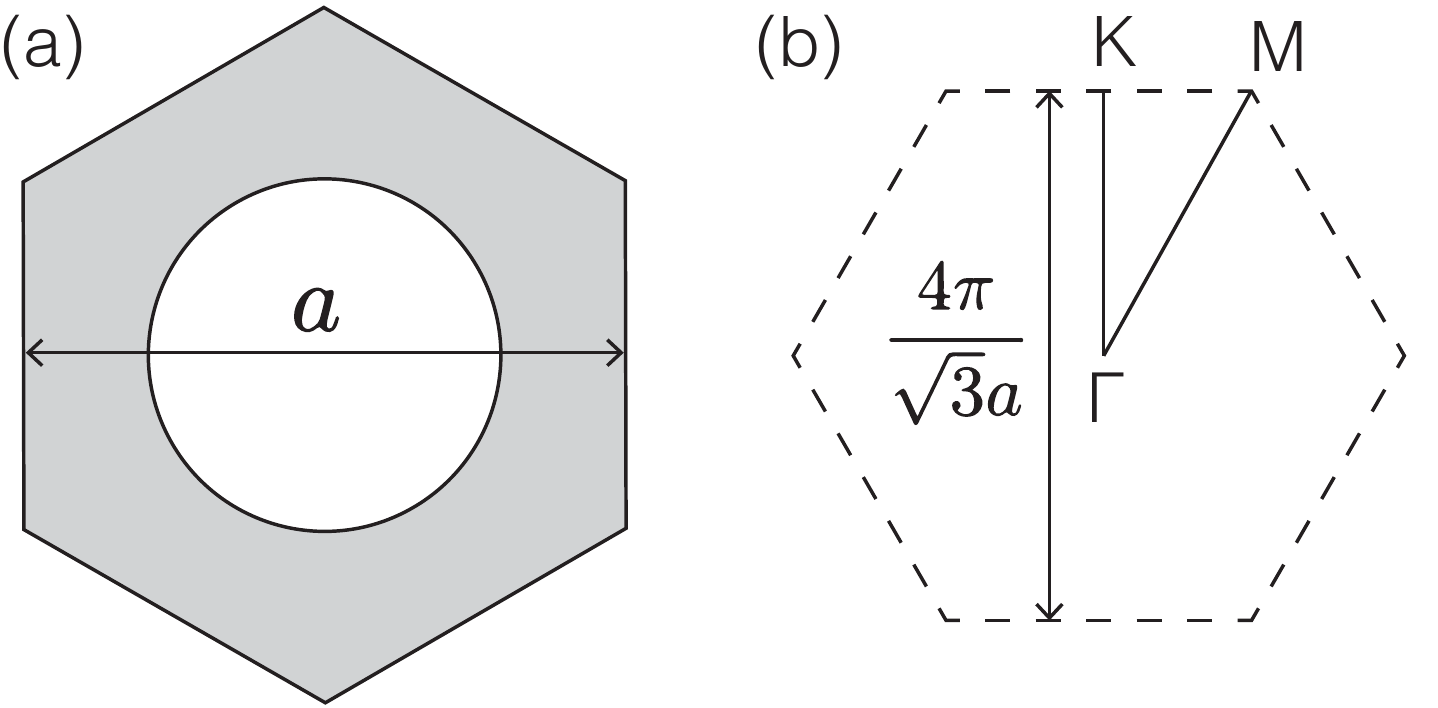}
 \caption{(a) Unit cell for hexagonal honeycomb lattice with periodicity $a$. (b) First Brillouin zone with symmetry points for the reduced wave vector.\label{fig:hexagon}}
\end{figure}We have used the finite element simulation (FEM) tool COMSOL Multiphysics\textsuperscript{\textregistered} to model highly confined third sound modes in a superfluid crystal lattice by solving the hydrodynamic equations for third sound:
The linearized Euler equation
\begin{equation}
\dot{\vec{v}}+\vec{v}\cdot\nabla\vec{v}=- \frac{3\alpha}{d^4} \vec{\nabla} \eta ,
\end{equation}
and the continuity equation for the film height
\begin{equation}
\dot{\eta}+\vec{v}\cdot \vec{\nabla}\eta=-d \,\vec{\nabla} \cdot \vec{v}.
\end{equation}
While the software does not include ready-made solvers for the third sound, one can map the equations to those for (first) sound in an ideal gas---as outlined in  Ref.\cite{Forstner2019}.
\begin{figure}[t]
\centering
\includegraphics[width=0.45\textwidth]{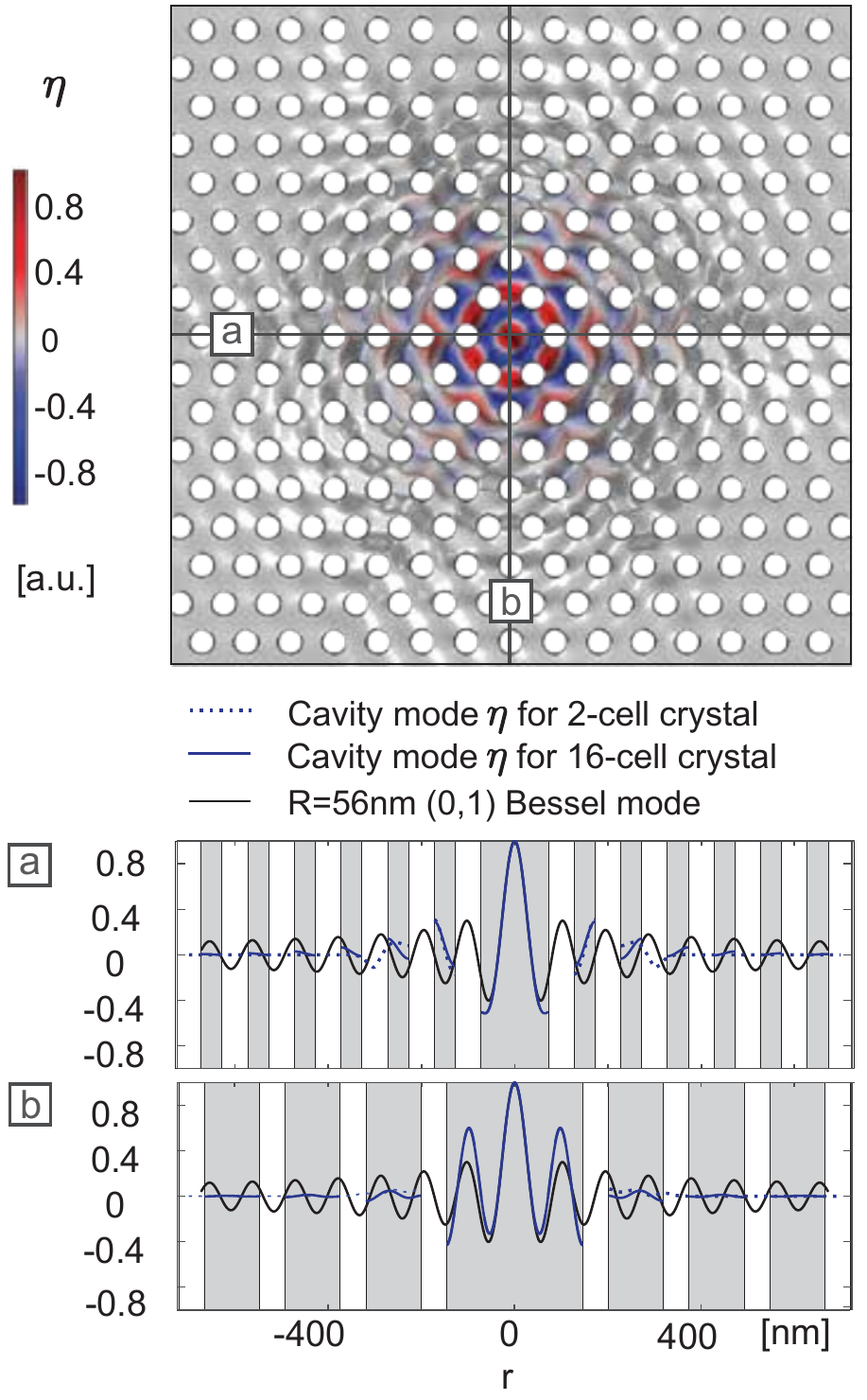}%
 \caption{Mode amplitude $\amp[r,\theta]$ for \SI{30}{\mega\hertz} third sound in a \SI{11}{\nano \metre} thick superfluid film condensed on a suspended silicon slab (with van der Waals coefficient $\avdw = \SI{3.5E-24}{m^5 . s^{-2}} $ \cite{PhysRevLett.30.1122}) perforated with \SI{55}{\nano \metre} diameter holes and $\SI{100}{\nano \metre}$ periodicity. Panels (a) and (b) show the horizontal and vertical cross sections respectively, in comparison to the \SI{31}{\mega\hertz} fundamental Bessel mode confined with free boundary conditions to a radius of \SI{56}{\nano \metre}. Finite element method simulations.\label{fig:crosssectionfig} 
}
\end{figure}
From the hexagonal unit cell on Fig. \ref{fig:hexagon} with lattice constant $a$, Floquet boundary conditions on the outer edges and free boundary conditions on the inner edge, the band structure is modeled by performing a stationary eigenfrequency analysis swept over the Floquet vector 
\begin{equation}
\vec{k}=\left( k_1 b_{1x}+k_2 b_{2x}, k_1 b_{1y}+k_2 b_{2y} \right)
\end{equation}
with $k_1$ and $k_2$ dimensionless parameters between 0 and 1 to cover the entire first Brillouin zone,
\begin{equation}
\vec{b_1}=\left( \frac{2\pi}{a}, \frac{-2\pi}{\sqrt{3}a} \right)
\end{equation}
the first reciprocal lattice vector, and
\begin{equation}
\vec{b_2}=\left(0, \frac{4\pi}{\sqrt{3}a} \right)
\end{equation}
the second reciprocal lattice vector.

For example then, the band structure for a $d=\SI{11}{\nano \metre}$ thick superfluid film on an $a=\SI{100}{\nano \metre}$ lattice with \SI{55}{\nano \metre} diameter holes is shown on Fig.6 in the main text and is found to have a band gap between 29 and \SI{30.5}{\mega\hertz}. By modeling the full crystal with a chosen amount of unit cells in either direction and the central hole removed to form a cavity, we obtained the mode amplitude $\amp[r,\theta]$ and the areal energy density 
\begin{equation}
u= \rhosf \alpha \left( \frac{-\amp}{d^3} +\frac{3\amp^2}{2 d^4} \right)
\end{equation}
for the \SI{30}{\mega\hertz} mode in the band gap. \begin{figure}[t]
\centering
\includegraphics[width=0.5\textwidth]{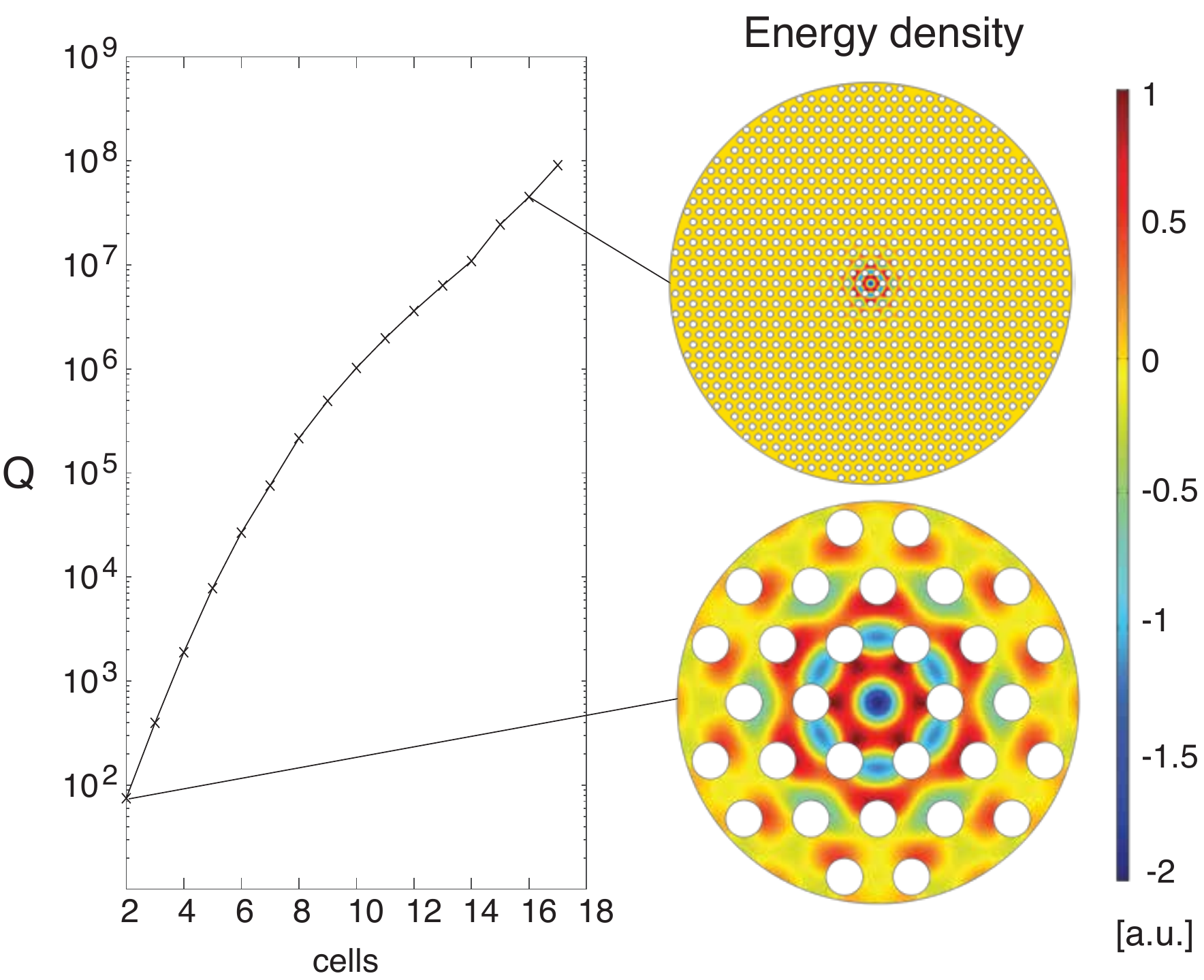}%
 \caption{Finite element model of the quality factor for a \SI{30}{\mega\hertz} third sound mode in \SI{11}{\nano \metre} thick superfluid film condensed on a suspended silicon slab perforated with \SI{55}{\nano \metre} diameter holes and crystal lattice constant $\SI{100}{\nano \metre}$.\label{fig:q_and_edens_fig}}
\end{figure}The amplitude $\amp$ is shown in Fig. \ref{fig:crosssectionfig} for a lattice comprising approximately 16 cells in either direction. It is compared to the amplitude found for a tiny crystal comprised of only two cells in either direction. It can be seen that while the mode profiles diverge slightly farther away from the central defect, they overlap in the central region. 

Since the cavity confinement in the crystal is not quite a perfectly circular confinement, it is instructive to look for the circular Bessel mode most closely resembling the profile of the trapped mode. For the present parameters, that is the $R=\SI{56}{\nano \metre}$ fundamental mode plotted in Fig. \ref{fig:crosssectionfig}. Its frequency is \SI{31}{\mega\hertz} (cf. \SI{30}{\mega\hertz} for the trapped mode). From the latter, we can estimate the intrinsic single-phonon nonlinear shift: here $\domo[\alef]/ 2\pi=\SI{35}{Hz}$. That means that a Q-factor $Q=\om / \Gamma$ in excess of $10^6$, or $Q \cdot f \geq \SI{6E13}{\hertz}$ is necessary to resolve the granular nature of the resonator. 

The energy density is plotted for the same two lattices on the right panel of Fig. \ref{fig:q_and_edens_fig}. The left panel shows the quality factor associated with acoustic radiation loss. As the number of cells in the lattice increases, the quality factor increases exponentially and exceeds $10^6$, the threshold for single-phonon resolution, for $10$-cell lattices.

\section{Critical velocity}
Liquid helium retains its superfluidity only as long as the particle velocity in the fluid $v$ remains below its critical velocity $v_c$, i.e. $\frac{v}{v_c} < 1$. Since this work is concerned with strong confinement of superfluid films, it is important to verify that superfluidity is retained for all radii $R$ and film thicknesses $d$ considered. From Ref. \cite{Atkins1959}, $v/v_{c} = \frac{\rhosf}{\rhohe} \frac{<x>}{d}$ \cite{Atkins1959}. 
\begin{figure}[H]
\centering
\includegraphics[width=0.39\textwidth]{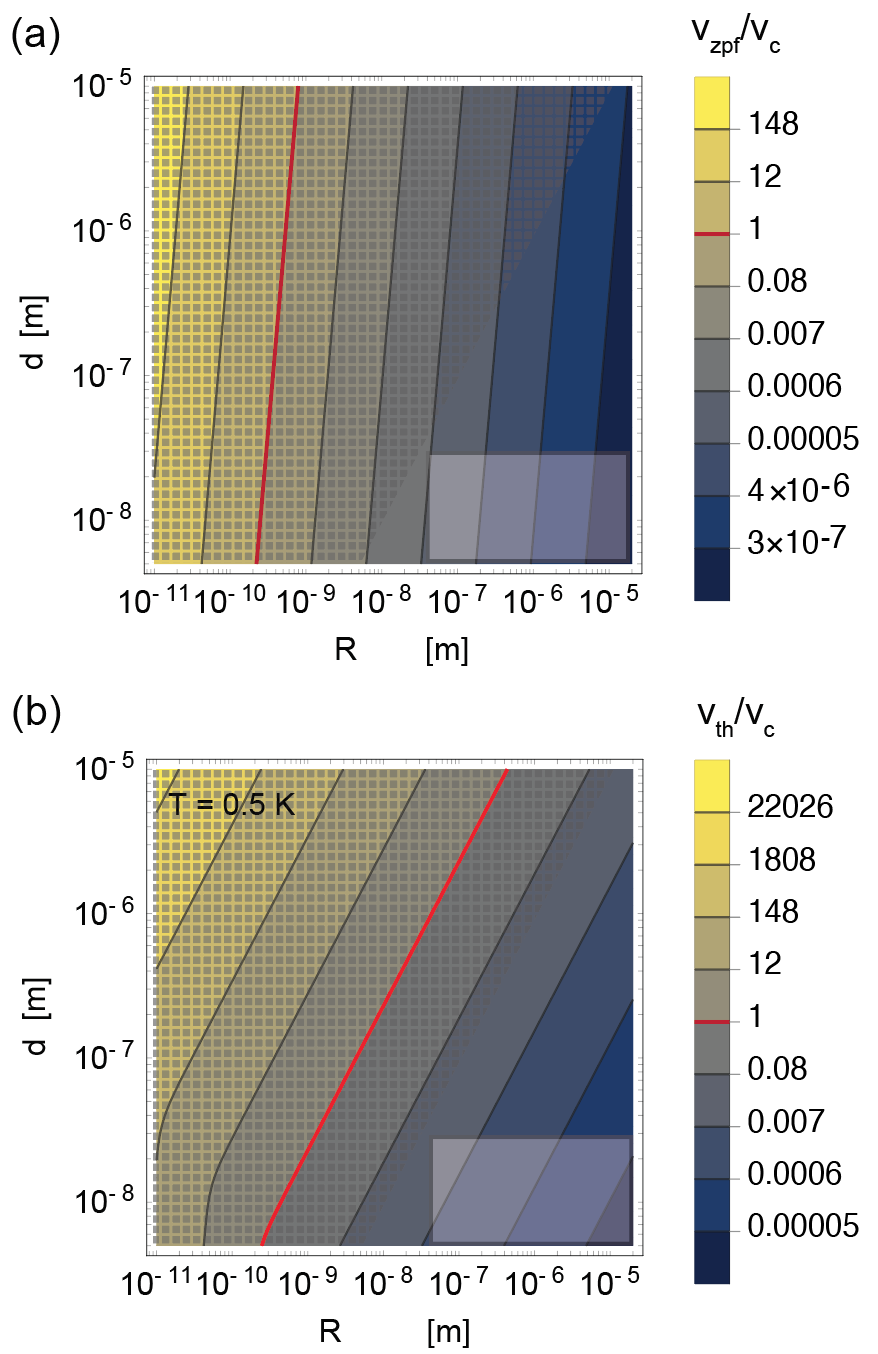}
 \caption{Ratio of superfluid particle velocity to critical velocity $v_c$. (a) Zero-point motion velocity $v_{\rm zpf}$ and (b) thermal motion velocity $v_{\rm th}$ at \SI{0.5}{\kelvin}. Shaded: regime of single-phonon nonlinear third-sound resonator. Red lines indicate $v_{zpf}=v_c$ and $v_{th}=v_c$ crossovers. Hatched: region outside of validity of this work ($d \geq R$)}\label{fig:vcrit}
\end{figure}
We consider two cases, first the case where the third sound mode is cooled to its motional ground state, and second the case where it is thermalized at temperature $T$. In the former case, the velocity ratio is
\begin{equation}
\frac{v_{\rm zpf}}{v_{c}} = \frac{\rhosf}{\rhohe} \frac{\xzpf[R,d]}{d}, \label{Aasd}
\end{equation}
where $v_{\rm zpf}$ is the zero point velocity, while in the latter case it is  
\begin{equation}
\frac{v_{\rm th}}{v_{c}} = \frac{\rhosf}{\rhohe}  \frac{
\xzpf[R,d] 
}{d} \sqrt{1+\frac{2}{e^{\hbar \om[R,d] /k_B T}-1}}. \label{Basd}
\end{equation}

In Figure~\ref{fig:vcrit} we show these two ratios as a function of the radius of the third sound mode  and the superfluid film thickness, taking $T=0.5$~K for the thermalized mode. It can be seen from these figures that 
the regime of operation we consider in the main text for nonlinear superfluid resonators (shaded) lies far outside of the regime where superfluidity breaks down, both for a third-sound resonator in its ground state (Fig. \ref{fig:vcrit}a) and for one cooled to \SI{0.5}{\kelvin} (Fig. \ref{fig:vcrit}b). Even in the most extreme parameters probing the limits of our model ($T=0.5$~K, $R=20$~nm, $d=5$~nm), the particle velocity remains two orders of magnitude lower than the critical velocity ($v_{c}$ = 250 $v_{th}$).

We finally note that, while it might be expected that the particle velocity would increase with decreasing film thickness from the explicit inverse-$d$ dependence of both Eqs.~(\ref{Aasd})~and~(\ref{Basd}), in fact the opposite is predicted. This is 
 due to the dependence of the resonator's spring constants and zero-point motion on the film thickness $d$, as derived in Eq. (9-11) and (23) in the main text.


\bibliographystyle{ieeetr}

\bibliography{nonlinsfbibliography_SI}{}